\RequirePackage[2020-02-02]{latexrelease}

\documentclass[%
 reprint,
 amsmath,amssymb,
 aps,
]{revtex4-2}
\usepackage{float}
\usepackage{graphicx}
\usepackage{dcolumn}
\usepackage{bm}
\usepackage{subcaption,lipsum}
\usepackage[margin=1.5cm]{geometry}

\begin{document}

\preprint{APS/123-QED}

\title{Interpretation of recently discovered single bottom baryons in the relativistic flux tube model}
\author{Pooja Jakhad}
\author{Juhi Oudichhya}
\author{Ajay Kumar Rai}
\affiliation{Department of Physics, Sardar Vallabhbhai National Institute of Technology, Surat, Gujarat-395007, India}
\date{\today}
\begin{abstract}
Following recent experimental progress in the study of bottom baryons, we systematically calculate the mass spectra of $\Lambda_{b}$, $\Xi_{b}$, $\Sigma_{b}$, $\Xi_{b}^{'}$, and $\Omega_{b}$ baryons with a quark-diquark picture in the framework of a relativistic flux tube model with spin-dependent interactions in the j-j coupling scheme.
Furthermore, we calculate the strong decay width of bottom baryons decaying into a bottom baryon and a light pseudoscalar meson.
A good agreement is found between the calculated masses and the experimentally available masses of singly bottom baryons.
By analysing both mass spectra and strong decay widths, we interpret $\Sigma_{b}(6097)$ as a $1P(3/2^{-})$ state and $\Xi_{b}(6100)$ as a $1P(1/2^{-})$ state of $\Xi_{b}$ baryon. The $\Xi_{b}(6227)$ is identified to be an orbital excitation $1P$ of the $\Xi_{b}^{'}$ baryon with $J^{P}=3/2^{-}$. Further, we determine $\Xi_{b}(6327)$ and $\Xi_{b}(6333)$ as a $1P(3/2^{-})$ state and $1P(5/2^{-})$ state, respectively, of $\Xi_{b}^{'}$ baryon.
From the obtained mass spectra, we construct the Regge trajectories in the $(J,M^{2})$ plane, which are found to be essentially linear, parallel, and equidistant. Our predictions for higher orbital and radial excited states can help experimentalists identify missing excited states of singly bottom baryons.

\end{abstract}
\maketitle


\section{INTRODUCTION}

\begin{figure*}
 	\includegraphics[scale=0.6]{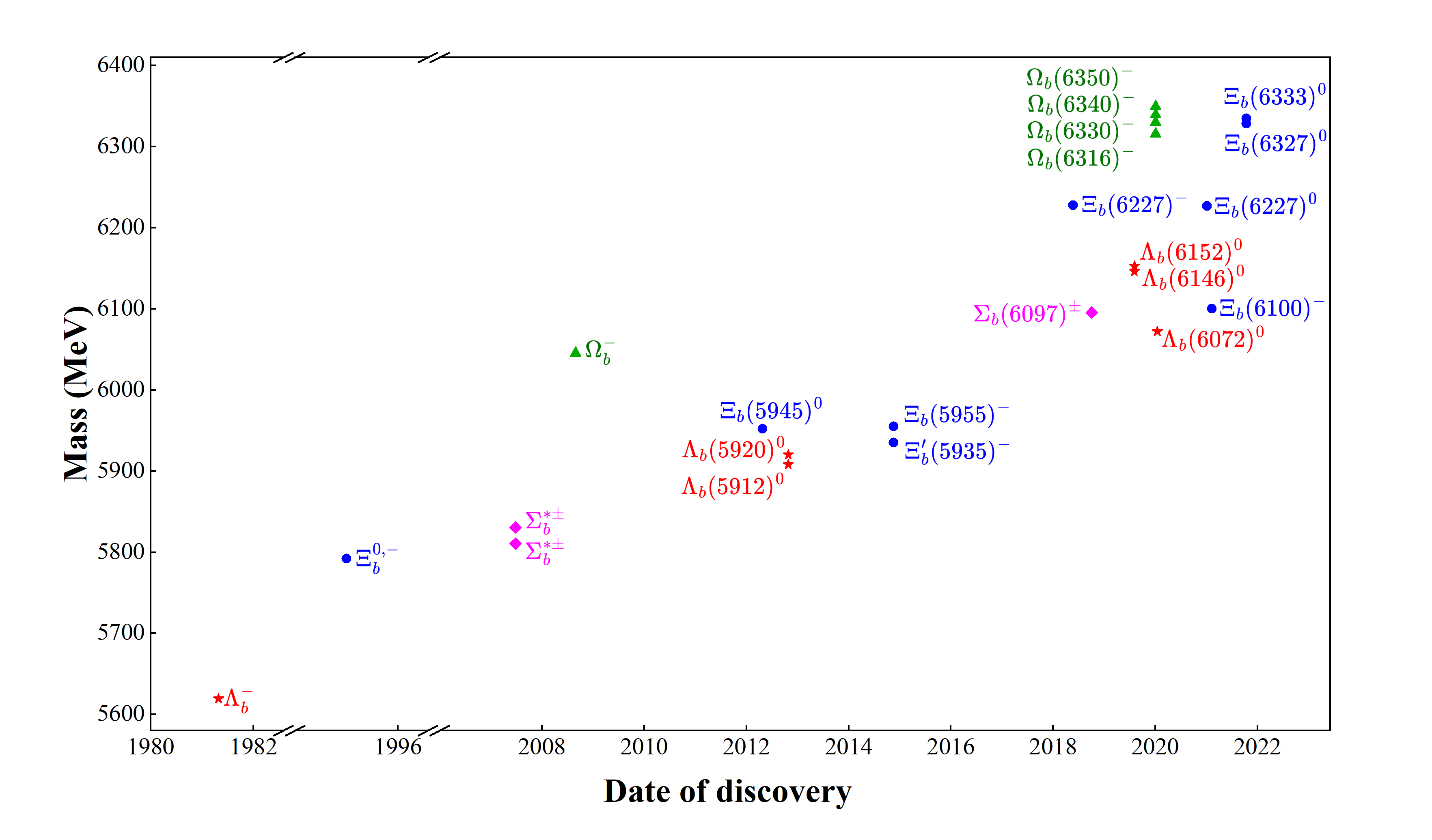}
	\caption{\label{fig:1}  The experimentally observed states of singly bottom baryons}
\end{figure*}

The discovery of a new hadron always comes up with a hint for us to better understand how quarks interact with each other in the hadronic system. Within the hadronic family, singly heavy baryons have an important position, as both chiral symmetry and heavy quark symmetry play a significant role in their dynamics. A thorough investigation of the observed singly heavy baryon can help us improve our understanding of the nature of the strong interaction in the domain of quark confinement. Experiments have observed more than 40 states of singly charmed baryons so far. However, searching for the single bottom baryon states is a difficult challenge for experimentalists since more energy and beam luminosity are required for their production. Furthermore, due to their short lifetime, they are extremely difficult to detect. Fig. \ref{fig:1} shows the experimental progress in discovering the singly bottom baryon states.

The first experimental observation of a singly bottom baryon, named $\Lambda_{b}(5620)^{0}$, was achieved in 1981 by CERN R415 \cite{Basile:1981wr}. After 14 years of this observation, in 1995, DELPHI announced the discovery of the first strange bottom baryons, $\Xi_{b}^{0}$ and $\Xi_{b}^{-}$ \cite{DELPHI:1995jet}. Then,  $\Sigma_{b}^{\pm}$ and $\Sigma_{b}^{*\pm}$, were reported by the CDF experiment in 2007, long after the previous discovery \cite{PhysRevLett.99.202001}. In the following year, the first doubly strange bottom baryon, $\Omega_{b}^{-}$, was also observed by the D0 detector at the Fermilab Tevatron Collider \cite{D0:2008sbw}. 

Then, in the following years, especially with the start of LHCb Run 1, in 2011, many bottom baryons were expected to be observed. In 2012, CMS collaboration came up with the observation of $\Xi_{b}(5945)^{0}$ state \cite{CMS:2012frl}.
In the same year, two narrow P-wave $\Lambda_{b}^{0}$ baryons, denoted as $\Lambda_{b}(5912)^{0}$ and $\Lambda_{b}(5920)^{0}$, were also discovered by the LHCb Collaboration \cite{PhysRevLett.109.172003}, which was later confirmed by the CDF Collaboration
 \cite{PhysRevD.88.072003}. In 2015, LHCb reported the ground state of $\Xi_{b}'$ baryon, denoted as $\Xi_{b}^{'}(5935)^{-}$, and the $\Xi_{b}(5955)^{-}$ state    \cite{LHCb:2014nae} by analysing the data of run 1. In 2016, the isospin partner of $\Xi_{b}(5955)^{-}$, which is denoted a $\Xi_{b}(5945)^{0}$, was confirmed by LHCb with more precise measurement of mass \cite{LHCb:2016mrc}.

At that point, many states were about to be identified as LHCb had taken collision data from run 2 in the period of 2015–2018. In 2018, they announced the discovery of $\Xi_{b}(6227)^{-}$ state \cite{PhysRevLett.121.072002} and two orbital excited states of $\Sigma_{b}$ baryon, $\Sigma_{b}(6097)^{+}$ and $\Sigma_{b}(6097)^{-}$ \cite{PhysRevLett.122.012001}. In 2019, they again reported two D-wave $\Lambda_{b}^{0}$ candidates, $\Lambda_{b}(6146)^{0}$ and $\Lambda_{b}(6152)^{0}$ \cite{PhysRevLett.123.152001}. The first observation of the four excited states of $\Omega_{b}$ baryon, named $\Omega_{b}(6316)^{-}$, $\Omega_{b}(6330)^{-}$, $\Omega_{b}(6340)^{-}$, and $\Omega_{b}(6350)^{-}$, was also announced by LHCb in 2020 \cite{LHCb:2020tqd}. Later in that year, they again observed $\Lambda_{b}(6070)^{0}$ state \cite{LHCb:2020lzx}, which was subsequently confirmed by the CMS \cite{2020135345} experiment. 

In 2021, the isospin partner of $\Xi_{b}(6227)^{-}$, named $\Xi_{b}(6227)^{0}$, is reported by LHCb \cite{LHCb:2020xpu}. In the same year, the CMS collaboration came up with the observation of $\Xi_{b}(6100)^{-}$ state \cite{CMS:2021rvl}. Later in the same year, two new $\Xi_{b}$ states, namely $\Xi_{b}(6327)^{0}$ and $\Xi_{b}(6333)^{0}$, are reported by the LHCb collaboration \cite{PhysRevLett.128.162001}.


These experimental discoveries have motivated a variety of theoretical studies. The systematic study of the mass spectra of all singly bottom baryons, up to high radial and orbital excited states, was first performed by Ebert \textit{et. al.} \cite{Ebert:2011kk}. The recent theoretical study on the mass spectra of singly bottom baryons includes the work by Garcia-Tecocoatzi \textit{et. al.} in which the Hamiltonian model is used with a three-quark and a quark-diquark picture of baryons \cite{Garcia-Tecocoatzi:2023btk}. The authors in ref. \cite{YU2023116183, li2022systematic} study the mass spectra of both strange and non-strange singly bottom baryons using the relativistic quark model. In addition to these recent studies, it has been studied by the non-relativistic constituent quark model \cite{PhysRevD.92.114029, doi:10.1142/S0217751X08041219, PhysRevD.102.014004, PhysRevD.104.054012}, the hyper-central constituent quark model \cite{Thakkar:2016dna,Kakadiya:2021jtv,Kakadiya:2022zvy}, the Regge trajectory model \cite{Pan:2023hwt}, the QCD spectral sum rules \cite{PhysRevD.94.114016, BAGAN1992176}, the QCD bag model \cite{HASENFRATZ1980401}, the QCD sum rule \cite{ Wang:2010it, PhysRevD.92.114007}, the lattice QCD \cite{PhysRevD.90.094004, Brown:2014ena}. The more theoretical studies with more references can be found in review articles \cite{Crede_2013, Chen_2017, HFLAV:2022esi, Chen_2023}. 

Although there are multiple theoretical and experimental approaches, very few states of single bottom baryons have been established. The spin-parity of $\Sigma_{b}(6097)$, $\Xi_{b}(6100)$ $\Xi_{b}(6227)$, $\Xi_{b}(6327)$, $\Xi_{b}(6333)$, $\Omega_{b}(6316)$, $\Omega_{b}(6330)$, $\Omega_{b}(6340)$, and $\Omega_{b}(6350)$ are still unknown. Assigning spin parity is crucial as it aids in determining their experimental properties. As various theoretical approaches yield different predictions about the spin parity for these states, it is important to do more theoretical investigations and compare them with experimental data in order to identify them. This motivates us to systematically examine the mass spectra of single-bottom baryons.

In Ref.\cite{article}, the authors have calculated the mass spectrum of $\Lambda_{c/b}$ and $\Xi_{c/b}$ baryons utilising a linear Regge relation derived in a relativistic flux tube model. However, they opt out of investigating other singly heavy baryonic systems ($\Sigma_{c/b}$, $\Xi_{c/b}^{'}$, and $\Omega_{c/b}$ baryons) containing vector diquarks due to the intricate nature of spin-dependent interactions.
In our previous work \cite{PhysRevD.108.014011,Jakhad:2024kkp}, we conducted calculations of the mass spectra for all singly charmed baryons using this linear Regge relation developed from the relativistic flux tube model that incorporates the spin-dependent interactions in the j-j coupling scheme. The aim of the present article is to extend this model to calculate the single-bottom baryon mass spectra. This will help us to assign possible spin parity to the experimentally detected states and to predict the masses of unobserved excited states, which can provide some significant information for future experiments.

The structure of the paper is as follows: In Section II, we describe the details of the relativistic flux tube model for singly bottom baryons, as well as the methodology employed to calculate their mass spectra. In section III,  we present the formulation to compute the strong decay widths. In Section IV, we discuss the results  and compare them with other theoretical estimations. In addition, we discuss  our assignment to the available experimental states  by examining their mass spectrum and decay widths. In Section V, we present our conclusion.

\section{MASS SPECTRUM}

\subsection{Singly bottom baryons in RFT model}

The singly bottom baryons can be seen as a bound system of a bottom quark ($b$) and two light quarks ($qq$, where q represents $u$, $d$, or $s$ quarks). There are different types of interactions in the system, such as quark-quark interaction, quark-gluon interaction, and gluon-gluon interaction, that make it a complex system to study.

To simplify this problem a bit, the heavy quark symmetry suggests that the coupling between two light quarks is stronger than the coupling between a bottom quark and a light quark \cite{PhysRevLett.66.1130}. It follows that two light quarks might couple first to form a diquark, which could then couple with a bottom quark, resulting in singly bottom baryonic states. In this way, we can reduce the three-body problem ($bqq$) into a two-body problem by taking a heavy-bottom-quark-light-diquark picture 
of singy bottom baryons. \textbf{In this picture, the diquark is assumed to stay in ground state i.e. the relative motion between two light quarks are restricted and the two light quarks excite together as a pair relative to the bottom quark. This mode of excitation is called $\lambda$-mode of excitation. In contrast, within the three-body picture of the baryon, there also exists the $\rho$-mode of excitation, which involves the relative motion between the two light quarks. In quark-diquark baryon picture, as the $\rho$-mode of excitation is absent, the number of possible states decreases significantly compare to that in three-body picture of the baryon \cite{Garcia-Tecocoatzi:2023btk,Ortiz-Pacheco:2023kjn}. As the $\rho$-mode excitations of singly bottom baryons are not observed experimentally, it supports the idea that singly bottom baryons are better described by a quark-diquark picture \cite{article}. Although $\rho$-mode excitations may not be observed for various reasons, particularly their higher energy levels, large decay widths, and suppressed transitions, this suggests that further experimental and theoretical studies are required to understand the structure of singly bottom baryons.}

For simplicity, we assume the quark-diquark baryon model in our investigation, which is also supported by a number of theoretical frameworks\cite{hooft2004minimal, PhysRevD.78.076003, narodetskii2008charm}. In this picture, the diquark and bottom quark are confined inside the baryon through strong interaction carried by gluons. One effective way to capture some essential features of confinement is the relativistic flux tube model. In this model, the confining interaction between bottom quark and diquark is carried out by a thin, string-like object called a flux tube. A gluonic field between the bottom quark and the diquark is restricted to a flux tube having a constant tension ($T$). The light quarks within diquark are assumed to stay in their ground state. The effect of interaction between two light quarks within diquark is included in the mass of diquark. The whole system of the bottom quark, diquark, and flux tube rotates around its centre of mass, giving rise to different quantum states of the system. A linear Regge relation between mass ($\bar{M}$) and angular momentum quantum number ($L$) of a singly heavy baryonic system can be obtained using this model as \cite{article,PhysRevD.108.014011}
	\begin{equation}
	\label{eq:1}
	(\bar{M}-m_b)^2=\frac{\sigma}{2}L+(m_{\mathcal{D}}+m_b v_{2}^{2}),
	\end{equation}
	Here, $\sigma=2 \pi T$.
As the diquark in its ground state with current mass $m_{1}$ and the bottom quark with current quark mass $m_{2}$ rotate with speeds $v_{1}$ and $v_{2}$, respectively, their effective masses are $m_{\mathcal{D}}=m_{1}/\sqrt{1-{v_{1}^{2}}}$ and $m_{b}=m_{2}/\sqrt{1-{v_{2}^{2}}}$, respectively.
The light diquark is assumed to rotate with ultra-relativistic speed, which leads to the assumption that it's speed, $v_{1}=1$.

The distance between the bottom quark and the diquark in this model is given as \cite{article}
\begin{equation}
\label{eq:2}
r=(v_1+v_2)\sqrt{\frac{8L}{\sigma}}.
\end{equation}

For a two-body picture of a heavy-light hadronic system, the quantum solution of the RFT model predicts that the Regge trajectories in the $(L,(\bar{M}-m_b)^2)$ plane, for different radial excitations, are parallel and equidistant to each other \cite{PhysRevD.49.4675, olsson1994relativistic, PhysRevD.60.074026}. This study leads us to modify our semi-classical relations (\ref{eq:1}) and (\ref{eq:2}) by replacing $ L $ with $ \lambda n_{r}+L $ (here, $n_{r}=n-1$, whereas n is the principle quantum number having values 1, 2, 3, etc., representing different radial excitations) to get a parallel and equidistant radial Regge trajectories in the $(L,(\bar{M}-m_b)^2)$ plane. The modified relationships are

\begin{equation}
\label{eq:3}
(\bar{M}-m_b)^2=\frac{\sigma}{2}[\lambda n_{r}+L]+(m_{\mathcal{D}}+m_b v_{2}^{2}),
\end{equation}

and
\begin{equation}
\label{eq:4}
r=(v_1+v_2)\sqrt{\frac{8[\lambda n_{r}+L]}{\sigma}}.
\end{equation}
\textbf{Here, $\lambda$ is a parameter of our model that defines the vertical distance between the Regge trajectories (corresponding to principle quantum numbers $n=1, 2, 3,..$) in the $(L,(\bar{M}-m_b)^2)$ plane.}

\subsection{Spin-dependent splittings and singly bottom baryon states }

Since the RFT model assumes the quarks to be spinless particles, we must now account for the contribution to mass from spin-dependent interactions from QCD motivated quark potential model, as  
\begin{equation}
\label{eq:14}
\Delta{M}= H_{so}+ H_{t}+ H_{ss} .
\end{equation} 
Here, $H_{so}$ is a spin-orbit interaction term, given as \cite{PhysRevD.105.074014}
\begin{equation} 
\label{eq:15}
\begin{split}
H_{so}&=  [(\frac{2\alpha}{3r^3}-\frac{b'}{2r}) \frac{1}{m_{\mathcal{D}}^2 }+\frac{4\alpha}{3r^3} \frac{1}{m_{\mathcal{D}} m_b }]\mathbf{L}\cdot \mathbf{S_{\mathcal{D}}}\\
& \hspace{0.3cm} +[(\frac{2\alpha}{3r^3}-\frac{b'}{2r}) \frac{1}{m_b^2 }+\frac{4\alpha}{3r^3} \frac{1}{ m_{\mathcal{D}} m_b}]\mathbf{L}\cdot \mathbf{S_b}.\\
\end{split}
\end{equation} 
It comes from the short-range one-gluon exchange contribution and the long-range Thomas-precession term. The spin of bottom quark and diquark is represented by $ \mathbf{S_b} $ and $ \mathbf{S_{\mathcal{D}}} $, respectively. $\mathbf{L}$ denotes the orbital angular momentum of the system. Further, the tensor interaction term,
\begin{equation} 
\label{eq:16}
\begin{split}
H_t&=\frac{4\alpha}{3r^3} \frac{1}{m_{\mathcal{D}} m_b} [\frac{3(\mathbf{S_{\mathcal{D}}}\cdot \mathbf{r})(\mathbf{S_b}\cdot \mathbf{r})}{r^2} -\mathbf{S_{\mathcal{D}}}\cdot \mathbf{S_b}].\\
\end{split}
\end{equation} 
results from magnetic-dipole-magnetic-dipole color hyperfine interaction. For simplicity we define, $ \mathbf{\hat{B}} $=$ 3(\mathbf{S_{\mathcal{D}}}\cdot \mathbf{r})(\mathbf{S_b}\cdot \mathbf{r})/r^2 -\mathbf{S_{\mathcal{D}}}\cdot \mathbf{S_b} $.  Lastly, the spin-spin contact hyperfine interaction is given as 
\begin{equation}
\label{eq:17}
\begin{split}
H_{ss}&=\frac{32\alpha\sigma_0^3}{9\sqrt{\pi} m_{\mathcal{D}} m_b}e^{-\sigma_0^2 r^2 } \mathbf{S_{\mathcal{D}}}\cdot \mathbf{S_b}.\\
\end{split}
\end{equation}
We can determine the parameters $b'$ and $\sigma_0$ using experimental data. Due to these spin-dependent interactions ($\Delta M$), the states with mass $\bar{M}$ will split into different states having mass $\bar{M}+\Delta M$.

\begin{figure}[h]
\includegraphics[scale=0.33]{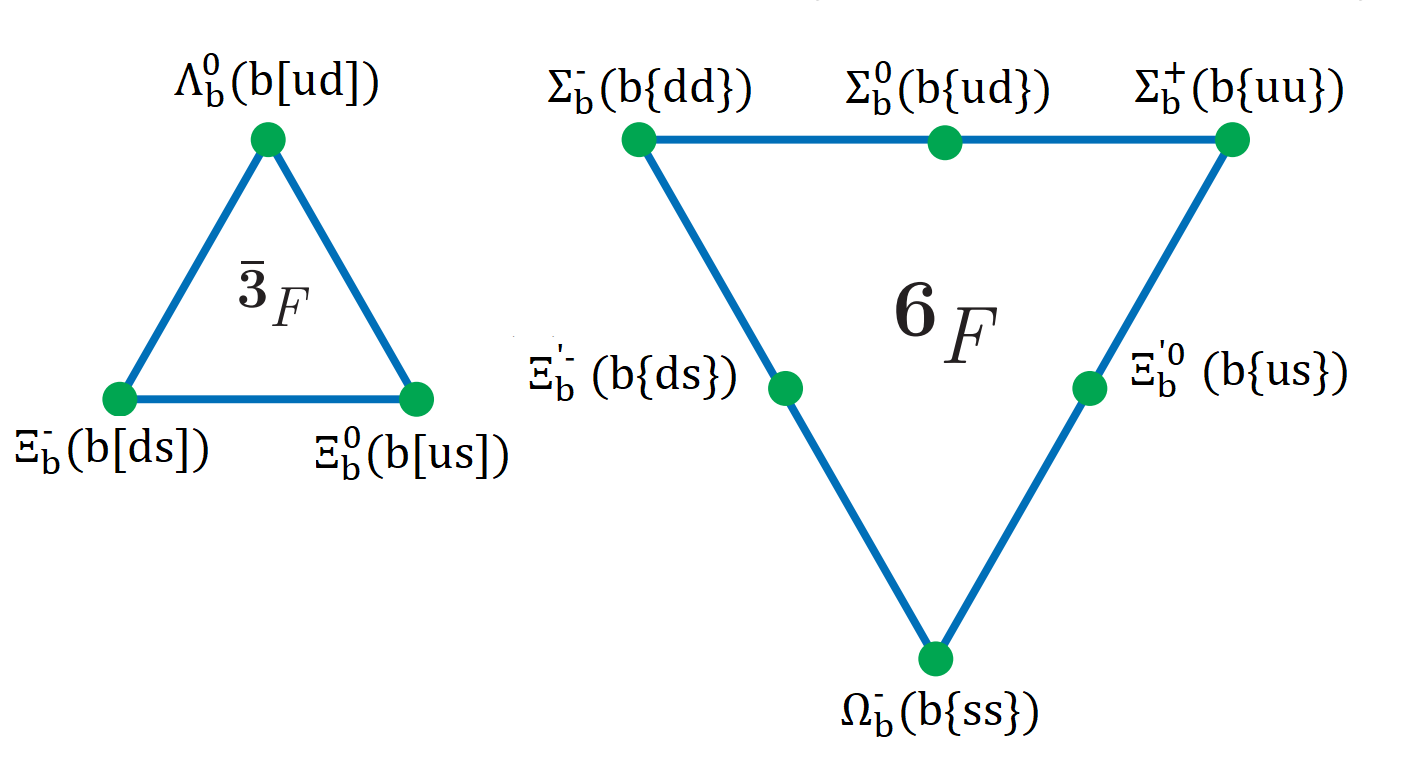}
\caption{\label{fig:2}  SU(3) flavor multiplets of singly bottom baryons. Anti-triplet($ \bar{3}_{F} $) consist of $\Lambda_b$ and $\Xi_b$ baryons, while sextet($ 6_{F} $) consist of $\Sigma_b$, $\Xi_b^{'}$, and $\Omega_b$ baryons.}
\end{figure}

If $SU_F(3)$ symmetry is considered for light quarks (u, d, and s), Pauli's exclusion principle states that the total wavefunction of a diquark is antisymmetric. The total wave function of diquark consists of a product of its space-, color-, flavor-, and spin-wave functions. As two light quarks are in their ground state, the space wave function of a diquark is symmetric. The color wave function of a diquark is always antisymmetric. These conditions restrict the product of the flavor- and spin-wave functions of diquark to being symmetric. As shown in Fig.\ref{fig:2}, the SU(3) flavor symmetry of light quarks arranges the singly bottom baryons into two groups: the first is antitriplet ($ \bar{3}_{F} $) with antisymmetric flavor wave function of light quarks, and the second is sextet ($ 6_F $) with symmetric flavor wave function of light quarks. To make the product of the flavor- and spin-wave functions of diquarks symmetric, diquarks belonging to the antitriplet and sextet flavor structures have to be spin-antisymmetric and spin-symmetric, respectively. This implies that the spin of a diquark present in $\Lambda_b$ and $\Xi_b$ baryons is $S_{\mathcal{D}}=0$, while that in $\Sigma_b$, $\Xi_b^{'}$, and $\Omega_b$ baryons is $S_{\mathcal{D}}=1$. We represent scalar diquarks with $S_{\mathcal{D}}=0$ by $[qq]$, while vector diquarks with $S_{\mathcal{D}}=1$ by $\{qq\}$.

For $\Lambda_{b}$ and $\Xi_{b}$ baryons, the spin-depentent interactions are simple as $S_{\mathcal{D}}=0$. $\mathbf{S_{b}}$ directly couple with  $\mathbf{L}$ to give  $\mathbf{J}=\mathbf{L}+\mathbf{S_b}$. Squaring this, we obtain the expectation value of $ \mathbf{L}\cdot \mathbf{S_b} $ as 
\begin{equation}
	\langle\mathbf{L}\cdot \mathbf{S_b}\rangle=\frac{1}{2}[J(J+1)-L(L+1)-S_b(S_b+1)].
\end{equation}
In spin-dependent interactions, only a term proportional to $\mathbf{L}\cdot \mathbf{S_b}$ in spin-orbit interaction ($H_{so}$) survives. This term splits the state with given values of $L$ into two different states having $J=L\pm1/2$ as listed in Tables \ref{tab:table2} and \ref{tab:table3}.

For $\Sigma_b$, $\Xi_b^{'}$, and $\Omega_b$ baryons, $\mathbf{S_{\mathcal{D}}}$ can couple with $\mathbf{S_b}$ and $\mathbf{L}$ in two ways. One possibility is that $\mathbf{S_{\mathcal{D}}}$ first couple with $\mathbf{S_b}$ to the total spin $\mathbf{S}=\mathbf{S_{\mathcal{D}}}+\mathbf{S_b}$, which subsequently couple with $\mathbf{L}$ giving total angular momentum $\mathbf{J}=\mathbf{L}+\mathbf{S}$. This way is called the  $L-S$  coupling scheme. 
The other way is that $\mathbf{S_{\mathcal{D}}}$ first couple with $\mathbf{L}$ to give the total angular momentum of diquark $\mathbf{j}=\mathbf{S_{\mathcal{D}}}+\mathbf{L}$, and then $\mathbf{j}$ couple with $\mathbf{S_{b}}$ to give $\mathbf{J}=\mathbf{j}+\mathbf{S_{b}}$. This scheme is known as the $j-j$ coupling scheme. The $j-j$ coupling scheme is preferred for singly bottom baryons due to their adherence to heavy quark symmetry. As a result of these couplings, the state with orbital angular momentum $L$ split into different states (as listed in Table \ref{tab:table1}, \ref{tab:table4}-\ref{tab:table6}) defined by $j$ and $J$, where $j$ is diquark's angular momentum quantum number and $J$ is the total angular momentum quantum number of the given state.
Accordingly, we calculate the expectation values of  $\mathbf{L}\cdot \mathbf{S_{\mathcal{D}}}$, $\mathbf{L}\cdot \mathbf{S_{b}}$, $\mathbf{S_{\mathcal{D}}}\cdot \mathbf{S_{b}}$, and $\mathbf{\hat{B}}$ in the $j-j$ coupling scheme and list them in Table \ref{tab:table1}. The computation of expectation values for these operators for $S$-wave, $P$-wave, and $D$-wave is presented in our previous work \cite{PhysRevD.108.014011}, and that for $F$-wave and $G$-wave is given in Appendix. 

\subsection{Determination of parameters}
This section lists the values of parameters for readers' convenience. The parameters involved in this theoretical model are $ m_b $, $ m_{\mathcal{D}} $, $ v_2 $, $\sigma$, $\lambda$, $\alpha$, $b'$, and $\sigma_{0}$. Some of these parameters have already been extracted in our previous work \cite{PhysRevD.108.014011} using experimentally available states of singly charmed baryons. 
The masses of diquarks ($m_{\mathcal{D}}$), with specific spin and quark combination ($[qq]$ or $\{qq\}$), were determined to be $ m_{\mathcal{D}[u,d]}=0.503 $ GeV, $ m_{\mathcal{D}[d,s]}=0.687 $ GeV, $m_{\mathcal{D}\{u,u\}}=0.714$ GeV, $ m_{\mathcal{D}\{d,s\}}=0.841$ GeV, and $ m_{\mathcal{D}\{s,s\}}=0.959$ GeV.
Parameters involved in spin-dependent relations were calculated to be $\alpha$=0.426, $b'$=-0.076 GeV$^{2}$, and $\sigma_{0}$=0.373248 GeV. The calculated value of $\lambda$ for $\Lambda_{b}$ and $\Xi_{b}$ baryons was 1.565, and that for $\Sigma_{b}$, $\Xi_{b}^{'}$, and $\Omega_{b}$ baryons was 1.295. We adopt these previously extracted parameters to calculate the mass spectra of single-bottom baryons so that consistency is maintained in the model.

Additionally, we use the current quark mass of the bottom quark ($m_{2}=4.18$ GeV \cite{Workman:2022ynf}) and the experimental mass of the ground state ($|1S,1/2^+ \rangle$) of the $\Lambda_{b}$ baryon to get $m_{b}=4.499$ GeV and $v_{2}=0.37$. We calculate the spin average mass of the $1P$-wave for the $\Lambda_{b}$ baryon from the experimentally available masses of $\Lambda_{b}(5912)^{0}$ and $\Lambda_{b}(5920)^{0}$ states to extract $\sigma_{\Lambda_{b}}=1.512$ GeV$^{2}$. However, for other singly bottom baryons, the states belonging to the $1P$-wave are yet to be identified, which restricts us from directly extracting $\sigma$ from experimental data. Within the singly bottom baryonic family, all systems have the same heavy component, which is a bottom quark, but the mass of the light diquark varies due to different quark combinations and its spin (0 or 1). Consequently, the string tension of these systems could be a function of the diquark's mass. In our previous work, for the singly charmed baryonic family, the string tension was assumed to be proportional to the $q^{th}$ power of the mass of diquark, i.e.
\begin{equation}
\label{eq:3.1}
\sigma\propto m_{\mathcal{D}}^{q} .
\end{equation} 
We were able to describe the experimentally observed excited states of singly charmed baryons based on this assumption. The value of $q$ was determined to be 0.661. Inspired by this, we assume that relation [\ref{eq:3.1}] can also be applied to the singly bottom baryonic family. The ratio of $\sigma$ for $\Xi_{b}$ baryon and that for $\Lambda_{b}$ baryon,
\begin{equation}
\label{eq:3.2}
\frac{\sigma_{\Xi_{b}}}{\sigma_{\Lambda_{b}}}=\left(\frac{m_{\mathcal{D}[d,s]}}{m_{\mathcal{D}[u,d]}}\right)^q,
\end{equation} 
allow us to find $\sigma_{\Xi_{b}}=1.857 $ GeV$^{2}$. Similarly, we obtain $\sigma_{\Sigma_{b}}=1.904$ GeV$^{2}$, $\sigma_{\Xi_{b}^{'}}=2.122$ GeV$^{2}$, and $\sigma_{\Omega_{b}}=2.315$ GeV$^{2}$. Once the parameters of this model are extracted, we calculate the masses of possible states of singly bottom baryons in the quark-diquark picture.

\begin{table}
\caption{\label{tab:table1}Expectation value of operators involved in spin-dependent interactions in the $j-j$ coupling scheme for possible states of singly heavy baryons having vector diquark \cite{PhysRevD.108.014011}.}
\begin{ruledtabular}
\begin{tabular}{cccccccccccc}

$(L, J, j)$	&	$\langle \mathbf{S_{\mathcal{D}}\cdot  L}\rangle$	&	$\langle \mathbf{S_{b}\cdot  L}\rangle$	&	$\langle\mathbf{\hat{B}}\rangle$	&	$\langle \mathbf{S_{\mathcal{D}}\cdot  S_{b}}\rangle$	\\ 
\hline
$(S, 1/2, 1 )$ 	&	0	&	0	&	0	&	-1	\\ 
$(S, 3/2, 1 )$ 	&	0	&	0	&	0	&	$ {1 }/{2}$	\\ 
$(P, 1/2, 0 )$ 	&	-2	&	0	&	0	&	0	\\ 
$(P, 1/2, 1 )$ 	&	-1	&	$-  {1 }/{2}$	&	-1	&	$-  {1 }/{2}$	\\ 
$(P, 3/2, 1 )$ 	&	-1	&	$ {1 }/{4}$	&	$  {1 }/{2}$	&	$ {1 }/{4}$	\\ 
$(P, 3/2, 2 )$ 	&	1	&	$-  {3 }/{4}$	&	$ {3 }/{10}$	&	$-  {3 }/{4}$	\\ 
$(P, 5/2, 2 )$ 	&	1	&	$  {1 }/{2}$	&	$-  {1 }/{5}$	&	$  {1 }/{2}$	\\ 
$(D, 1/2, 1 )$ 	&	-3	&	$-  {3 }/{2}$	&	-1	&	$  {1 }/{2}$	\\ 
$(D, 3/2, 1 )$ 	&	-3	&	$  {3 }/{4}$	&	$  {1 }/{2}$	&	$- {1 }/{4}$	\\ 
$(D, 3/2, 2 )$ 	&	-1	&	$-  {5 }/{4}$	&	$-  {1 }/{2}$	&	$-  {1 }/{4}$	\\ 
$(D, 5/2, 2 )$ 	&	2	&	$-  {4 }/{3}$	&	$ {8 }/{21}$	&	$-  {2 }/{3}$	\\ 
$(D, 5/2, 3 )$ 	&	-1	&	$  {5 }/{6}$	&	$  {1 }/{3}$	&	$  {1 }/{6}$	\\ 
$(D, 7/2, 3 )$ 	&	2	&	1	&	$-  {2 }/{7}$	&	$  {1 }/{2}$	\\ 
$(F, 3/2, 2 )$ 	&	-4	&	-2	&	$-  {4 }/{5}$	&	$  {1 }/{2}$	\\ 
$(F, 5/2, 2 )$ 	&	-4	&	$  {4 }/{3}$	&	$  {8 }/{15}$	&	$ - {1 }/{3}$	\\ 
$(F, 5/2, 3 )$ 	&	-1	&	$ - {11 }/{6}$	&	$ - {1 }/{3}$	&	$ - {1 }/{6}$	\\ 
$(F, 7/2, 3 )$ 	&	3	&	$ - {15 }/{8}$	&	$  {5 }/{12}$	&	$ - {5 }/{8}$	\\ 
$(F, 7/2, 4 )$ 	&	-1	&	$  {11 }/{8}$	&	$  {1 }/{4}$	&	$  {1 }/{8}$	\\ 
$(F, 9/2, 4 )$ 	&	3	&	$  {3 }/{2}$	&	$ - {1 }/{3}$	&	$  {1 }/{2}$	\\ 
$(G, 5/2, 3 )$ 	&	-5	&	$ - {5 }/{2}$	&	$ - {5 }/{7}$	&	$  {1 }/{2}$	\\ 
$(G, 7/2, 3 )$ 	&	-5	&	$  {15 }/{8}$	&	$  {15 }/{28}$	&	$ - {3 }/{8}$	\\ 
$(G, 7/2, 4 )$ 	&	-1	&	$ - {19 }/{8}$	&	$ - {1 }/{4}$	&	$ - {1 }/{8}$	\\ 
$(G, 9/2, 4 )$ 	&	4	&	$ - {12 }/{5}$	&	$  {24 }/{55}$	&	$ - {3 }/{5}$\\ 
$(G, 9/2, 5 )$ 	&	-1	&	$  {19 }/{10}$	&	$  {1 }/{5}$	&	$  {1 }/{10}$	\\ 
$(G, 11/2, 5 )$ 	&	4	&	2	&	$ - {4 }/{11}$	&	$  {1 }/{2}$	\\ 

\end{tabular}
\end{ruledtabular}
\end{table}  
\begin{table*}
\caption{\label{tab:table2} Masses of $\Lambda_{b}$ baryonic states predicted in the present work with the masses from experiments (PDG) and other theoretical studies. The masses are expressed in units of MeV. The asterisk (*) denotes that these experimental masses are taken as inputs to determine parameters.
}
\begin{ruledtabular}
\begin{tabular}{cccccccccccc}
$ (n, L, J, j) $	&	States\textit{$|nL,J^P\rangle$}	&	Present	&	PDG \cite{Workman:2022ynf}	&	\cite{Ebert:2011kk}	&	\cite{Garcia-Tecocoatzi:2023btk}	&	\cite{article}	&	\cite{YU2023116183}	&	\cite{Oudichhya:2021yln}	\\ 
\hline																	
(1, 0, 1/2, 0)	&	\textit{$|1S,1/2^+\rangle$}	&	5619.6	&	5619.60(0.17)*	&	5620	&	5611	&	5619	&	5622	&	5620	\\ 
(2, 0, 1/2, 0)	&	\textit{$|2S,1/2^+\rangle$}	&	6061.0	&	6072.30(2.90)	&	6089	&	6233	&		&	6041	&	6026	\\ 
(3, 0, 1/2, 0)	&	\textit{$|3S,1/2^+\rangle$}	&	6402.6	&		&	6455	&		&		&	6352	&	6406	\\ 
(4, 0, 1/2, 0)	&	\textit{$|4S,1/2^+\rangle$}	&	6691.6	&		&	6756	&		&		&	6388	&	6765	\\ 
(5, 0, 1/2, 0)	&	\textit{$|5S,1/2^+\rangle$}	&	6946.7	&		&	7015	&		&		&		&	7106	\\ 
(6, 0, 1/2, 0)	&	\textit{$|6S,1/2^+\rangle$}	&	7177.6	&		&	7256	&		&		&		&	7431	\\ 
(7, 0, 1/2, 0)	&	\textit{$|7S,1/2^+\rangle$}	&	7390.1	&		&		&		&		&		&		\\ 
(1, 1, 1/2, 1)	&	\textit{$|1P,1/2^-\rangle$}	&	5908.4	&	5912.19(0.17)*	&	5930	&	5916	&	5911	&	5898	&	5930	\\ 
(1, 1, 3/2, 1)	&	\textit{$|1P,3/2^-\rangle$}	&	5922.0	&	5920.09(0.17)*	&	5942	&	5925	&	5920	&	5913	&	5924	\\ 
(2, 1, 1/2, 1)	&	\textit{$|2P,1/2^-\rangle$}	&	6284.3	&		&	6326	&		&		&	6238	&		\\ 
(2, 1, 3/2, 1)	&	\textit{$|2P,3/2^-\rangle$}	&	6287.9	&		&	6333	&		&		&	6249	&	6304	\\ 
(3, 1, 1/2, 1)	&	\textit{$|3P,1/2^-\rangle$}	&	6590.5	&		&	6645	&		&		&	6544	&		\\ 
(3, 1, 3/2, 1)	&	\textit{$|3P,3/2^-\rangle$}	&	6592.4	&		&	6651	&		&		&	6552	&	6662	\\ 
(4, 1, 1/2, 1)	&	\textit{$|4P,1/2^-\rangle$}	&	6856.8	&		&	6917	&		&		&	6566	&		\\ 
(4, 1, 3/2, 1)	&	\textit{$|4P,3/2^-\rangle$}	&	6858.1	&		&	6922	&		&		&	6575	&	7002	\\ 
(5, 1, 1/2, 1)	&	\textit{$|5P,1/2^-\rangle$}	&	7095.9	&		&	7157	&		&		&		&		\\ 
(5, 1, 3/2, 1)	&	\textit{$|5P,3/2^-\rangle$}	&	7096.8	&		&	7171	&		&		&		&	7327	\\ 
(6, 1, 1/2, 1)	&	\textit{$|6P,1/2^-\rangle$}	&	7314.6	&		&		&		&		&		&		\\ 
(6, 1, 3/2, 1)	&	\textit{$|6P,3/2^-\rangle$}	&	7315.4	&		&		&		&		&		&		\\ 
(1, 2, 3/2, 2)	&	\textit{$|1D,3/2^+\rangle$}	&	6157.7	&	6146.20(0.40)	&	6190	&	6224	&	6147	&	6137	&	6128	\\ 
(1, 2, 5/2, 2)	&	\textit{$|1D,5/2^+\rangle$}	&	6166.2	&	6152.50(0.40)	&	6196	&	6239	&	6153	&	6145	&	6213	\\ 
(2, 2, 3/2, 2)	&	\textit{$|2D,3/2^+\rangle$}	&	6484.6	&		&	6526	&		&		&	6432	&		\\ 
(2, 2, 5/2, 2)	&	\textit{$|2D,5/2^+\rangle$}	&	6488.5	&		&	6531	&		&		&	6440	&	6527	\\ 
(3, 2, 3/2, 2)	&	\textit{$|3D,3/2^+\rangle$}	&	6763.7	&		&	6811	&		&		&	6705	&		\\ 
(3, 2, 5/2, 2)	&	\textit{$|3D,5/2^+\rangle$}	&	6766.2	&		&	6814	&		&		&	6709	&	6826	\\ 
(4, 2, 3/2, 2)	&	\textit{$|4D,3/2^+\rangle$}	&	7011.8	&		&	7060	&		&		&	6757	&		\\ 
(4, 2, 5/2, 2)	&	\textit{$|4D,5/2^+\rangle$}	&	7013.6	&		&	7063	&		&		&	6763	&	7113	\\ 
(5, 2, 3/2, 2)	&	\textit{$|5D,3/2^+\rangle$}	&	7237.3	&		&		&		&		&		&		\\ 
(5, 2, 5/2, 2)	&	\textit{$|5D,5/2^+\rangle$}	&	7238.8	&		&		&		&		&		&	7389	\\ 
(1, 3, 5/2, 3)	&	\textit{$|1F,5/2^-\rangle$}	&	6372.4	&		&	6408	&		&	6346	&	6338	&	6320	\\ 
(1, 3, 7/2, 3)	&	\textit{$|1F,7/2^-\rangle$}	&	6379.3	&		&	6411	&		&	6351	&	6343	&	6489	\\ 
(2, 3, 5/2, 3)	&	\textit{$|2F,5/2^-\rangle$}	&	6666.5	&		&	6705	&		&		&	6616	&		\\ 
(2, 3, 7/2, 3)	&	\textit{$|2F,7/2^-\rangle$}	&	6670.6	&		&	6708	&		&		&	6622	&		\\ 
(3, 3, 5/2, 3)	&	\textit{$|3F,5/2^-\rangle$}	&	6924.7	&		&	6964	&		&		&	6849	&		\\ 
(3, 3, 7/2, 3)	&	\textit{$|3F,7/2^-\rangle$}	&	6927.5	&		&	6966	&		&		&	6852	&		\\ 
(4, 3, 5/2, 3)	&	\textit{$|4F,5/2^-\rangle$}	&	7157.8	&		&	7196	&		&		&	6932	&		\\ 
(4, 3, 7/2, 3)	&	\textit{$|4F,7/2^-\rangle$}	&	7159.9	&		&	7197	&		&		&	6936	&		\\ 
(5, 3, 5/2, 3)	&	\textit{$|5F,5/2^-\rangle$}	&	7371.9	&		&		&		&		&		&		\\ 
(5, 3, 7/2, 3)	&	\textit{$|5F,7/2^-\rangle$}	&	7373.6	&		&		&		&		&		&		\\ 
(1, 4, 7/2, 4)	&	\textit{$|1G,7/2^+\rangle$}	&	6564.6	&		&	6598	&		&	6523	&	6514	&	6506	\\ 
(1, 4, 9/2, 4)	&	\textit{$|1G,9/2^+\rangle$}	&	6570.7	&		&	6599	&		&	6526	&	6517	&	6754	\\ 
(2, 4, 7/2, 4)	&	\textit{$|2G,7/2^+\rangle$}	&	6834.3	&		&	6867	&		&		&	6793	&		\\ 
(2, 4, 9/2, 4)	&	\textit{$|2G,9/2^+\rangle$}	&	6838.4	&		&	6868	&		&		&	6798	&		\\ 
(3, 4, 7/2, 4)	&	\textit{$|3G,7/2^+\rangle$}	&	7075.7	&		&		&		&		&	6986	&		\\ 
(3, 4, 9/2, 4)	&	\textit{$|3G,9/2^+\rangle$}	&	7078.7	&		&		&		&		&	6989	&		\\ 
(1, 5, 9/2, 5)	&	\textit{$|1H,9/2^-\rangle$}	&	6740.2	&		&	6767	&		&		&	7093	&	6687	\\ 
(1, 5,11/2, 5)	&	\hspace{1.55mm}\textit{$|1H,11/2^-\rangle$}	&	6745.8	&		&	6766	&		&		&	7095	&	7009	\\ 
(2, 5, 9/2, 5)	&	\textit{$|2H,9/2^-\rangle$}	&	6990.9	&		&		&		&		&		&		\\ 
(2, 5,11/2, 5)	&	\hspace{1.55mm}\textit{$|2H,11/2^-\rangle$}	&	6994.9	&		&		&		&		&		&		\\

\end{tabular}
\end{ruledtabular}
\end{table*}  

\begin{table*}
\caption{\label{tab:table3}
The same as Table \ref{tab:table2}, but with regard to the $\Xi_{b}$ baryonic states.}

\begin{ruledtabular}
\begin{tabular}{cccccccccccc}
$ (n, L, J, j) $	&	States\textit{$|nL,J^P\rangle$}	&	Present	&	PDG \cite{Workman:2022ynf}	&	\cite{Ebert:2011kk}	&	\cite{Garcia-Tecocoatzi:2023btk}	&	\cite{article}	&	\cite{li2022systematic}	&	\cite{Oudichhya:2021yln}	\\
\hline																	
(1, 0, 1/2, 0)	&	\textit{$|1S,1/2^+\rangle$}	&	5803.6	&	5797.0(0.6)	&	5803	&	5801	&	5801	&	5806	&	5792	\\ 
(2, 0, 1/2, 0)	&	\textit{$|2S,1/2^+\rangle$}	&	6275.7	&		&	6266	&	6377	&		&	6224	&	6203	\\ 
(3, 0, 1/2, 0)	&	\textit{$|3S,1/2^+\rangle$}	&	6646.3	&		&	6601	&		&		&	6480	&	6588	\\ 
(4, 0, 1/2, 0)	&	\textit{$|4S,1/2^+\rangle$}	&	6961.8	&		&	6913	&		&		&	6568	&	6952	\\ 
(5, 0, 1/2, 0)	&	\textit{$|5S,1/2^+\rangle$}	&	7241.2	&		&	7165	&		&		&		&	7298	\\ 
(6, 0, 1/2, 0)	&	\textit{$|6S,1/2^+\rangle$}	&	7494.7	&		&	7415	&		&		&		&	7629	\\ 
(7, 0, 1/2, 0)	&	\textit{$|7S,1/2^+\rangle$}	&	7728.3	&		&		&		&		&		&		\\ 
(1, 1, 1/2, 1)	&	\textit{$|1P,1/2^-\rangle$}	&	6111.8	&	6100.3(0.6)	&	6120	&	6082	&	6097	&	6084	&	6120	\\ 
(1, 1, 3/2, 1)	&	\textit{$|1P,3/2^-\rangle$}	&	6125.7	&		&	6130	&	6092	&	6106	&	6097	&	6093	\\ 
(2, 1, 1/2, 1)	&	\textit{$|2P,1/2^-\rangle$}	&	6517.8	&		&	6496	&		&		&	6421	&		\\ 
(2, 1, 3/2, 1)	&	\textit{$|2P,3/2^-\rangle$}	&	6521.5	&		&	6502	&		&		&	6432	&	6460	\\ 
(3, 1, 1/2, 1)	&	\textit{$|3P,1/2^-\rangle$}	&	6851.4	&		&	6805	&		&		&	6690	&		\\ 
(3, 1, 3/2, 1)	&	\textit{$|3P,3/2^-\rangle$}	&	6853.4	&		&	6810	&		&		&	6700	&	6807	\\ 
(4, 1, 1/2, 1)	&	\textit{$|4P,1/2^-\rangle$}	&	7142.8	&		&	7068	&		&		&	6732	&		\\ 
(4, 1, 3/2, 1)	&	\textit{$|4P,3/2^-\rangle$}	&	7144.1	&		&	7073	&		&		&	6739	&	7138	\\ 
(5, 1, 1/2, 1)	&	\textit{$|5P,1/2^-\rangle$}	&	7405.0	&		&	7302	&		&		&		&		\\ 
(5, 1, 3/2, 1)	&	\textit{$|5P,3/2^-\rangle$}	&	7406.0	&		&	7306	&		&		&		&	7453	\\ 
(6, 1, 1/2, 1)	&	\textit{$|6P,1/2^-\rangle$}	&	7645.4	&		&		&		&		&		&		\\ 
(6, 1, 3/2, 1)	&	\textit{$|6P,3/2^-\rangle$}	&	7646.2	&		&		&		&		&		&		\\ 
(1, 2, 3/2, 2)	&	\textit{$|1D,3/2^+\rangle$}	&	6380.6	&		&	6366	&	6368	&	6344	&	6320	&	6316	\\ 
(1, 2, 5/2, 2)	&	\textit{$|1D,5/2^+\rangle$}	&	6389.4	&		&	6373	&	6383	&	6349	&	6327	&	6380	\\ 
(2, 2, 3/2, 2)	&	\textit{$|2D,3/2^+\rangle$}	&	6735.9	&		&	6690	&		&		&	6613	&		\\ 
(2, 2, 5/2, 2)	&	\textit{$|2D,5/2^+\rangle$}	&	6740.0	&		&	6696	&		&		&	6621	&	6687	\\ 
(3, 2, 3/2, 2)	&	\textit{$|3D,3/2^+\rangle$}	&	7040.9	&		&	6966	&		&		&	6883	&		\\ 
(3, 2, 5/2, 2)	&	\textit{$|3D,5/2^+\rangle$}	&	7043.5	&		&	6970	&		&		&	6888	&	6980	\\ 
(4, 2, 3/2, 2)	&	\textit{$|4D,3/2^+\rangle$}	&	7312.7	&		&	7208	&		&		&	6890	&		\\ 
(4, 2, 5/2, 2)	&	\textit{$|4D,5/2^+\rangle$}	&	7314.6	&		&	7212	&		&		&	6894	&	7262	\\ 
(5, 2, 3/2, 2)	&	\textit{$|5D,3/2^+\rangle$}	&	7560.4	&		&		&		&		&		&		\\ 
(5, 2, 5/2, 2)	&	\textit{$|5D,5/2^+\rangle$}	&	7561.9	&		&		&		&		&		&	7533	\\ 
(1, 3, 5/2, 3)	&	\textit{$|1F,5/2^-\rangle$}	&	6613.7	&		&	6577	&		&	6555	&	6518	&	6506	\\ 
(1, 3, 7/2, 3)	&	\textit{$|1F,7/2^-\rangle$}	&	6620.8	&		&	6581	&		&	6559	&	6523	&	6654	\\ 
(2, 3, 5/2, 3)	&	\textit{$|2F,5/2^-\rangle$}	&	6934.6	&		&	6863	&		&		&	6795	&		\\ 
(2, 3, 7/2, 3)	&	\textit{$|2F,7/2^-\rangle$}	&	6938.7	&		&	6867	&		&		&	6801	&		\\ 
(3, 3, 5/2, 3)	&	\textit{$|3F,5/2^-\rangle$}	&	7217.3	&		&	7114	&		&		&	7032	&		\\ 
(3, 3, 7/2, 3)	&	\textit{$|3F,7/2^-\rangle$}	&	7220.2	&		&	7117	&		&		&	7034	&		\\ 
(4, 3, 5/2, 3)	&	\textit{$|4F,5/2^-\rangle$}	&	7473.0	&		&	7339	&		&		&	7057	&		\\ 
(4, 3, 7/2, 3)	&	\textit{$|4F,7/2^-\rangle$}	&	7475.3	&		&	7342	&		&		&	7060	&		\\ 
(5, 3, 5/2, 3)	&	\textit{$|5F,5/2^-\rangle$}	&	7708.4	&		&		&		&		&		&		\\ 
(5, 3, 7/2, 3)	&	\textit{$|5F,7/2^-\rangle$}	&	7710.2	&		&		&		&		&		&		\\ 
(1, 4, 7/2, 4)	&	\textit{$|1G,7/2^+\rangle$}	&	6823.2	&		&	6760	&		&	6743	&	6692	&	6690	\\ 
(1, 4, 9/2, 4)	&	\textit{$|1G,9/2^+\rangle$}	&	6829.5	&		&	6762	&		&	6747	&	6695	&	6918	\\ 
(2, 4, 7/2, 4)	&	\textit{$|2G,7/2^+\rangle$}	&	7118.2	&		&	7020	&		&		&	6970	&		\\ 
(2, 4, 9/2, 4)	&	\textit{$|2G,9/2^+\rangle$}	&	7122.4	&		&	7032	&		&		&	6975	&		\\ 
(3, 4, 7/2, 4)	&	\textit{$|3G,7/2^+\rangle$}	&	7382.9	&		&		&		&		&	7167	&		\\ 
(3, 4, 9/2, 4)	&	\textit{$|3G,9/2^+\rangle$}	&	7386.1	&		&		&		&		&	7169	&		\\ 
(1, 5, 9/2, 5)	&	\textit{$|1H,9/2^-\rangle$}	&	7015.2	&		&	6933	&		&		&		&		\\ 
(1, 5,11/2, 5)	&	\hspace{1.55mm}\textit{$|1H,11/2^-\rangle$}	&	7021.0	&		&	6934	&		&		&		&	7712	\\ 
(2, 5, 9/2, 5)	&	\textit{$|2H,9/2^-\rangle$}	&	7289.8	&		&		&		&		&		&		\\ 
(2, 5,11/2, 5)	&	\hspace{1.55mm}\textit{$|2H,11/2^-\rangle$}	&	7294.1	&		&		&		&		&		&		\\

\end{tabular}
\end{ruledtabular}
\end{table*} 

\begin{table*}
\caption{\label{tab:table4}
The same as Table \ref{tab:table2}, but with regard to the  $\Sigma_b$ baryonic states.}

\begin{ruledtabular}
\begin{tabular}{ccccccccccc}
$(n, L, J, j)$	&	States\textit{$|nL, J^P\rangle$}	&	Present	&	PDG \cite{Workman:2022ynf}	&	\cite{Ebert:2011kk}	&	\cite{Garcia-Tecocoatzi:2023btk}	&	\cite{YU2023116183}	&	\cite{Oudichhya:2021yln}	\\
\hline															
(1, 0, 1/2, 1)	&	\textit{$|1S, 1/2^+\rangle$}	&	5816.2	&	5810.56(0.25)	&	5808	&	5811	&	5820	&	5811	\\
(1, 0, 3/2, 1)	&	\textit{$|1S, 3/2^+\rangle$}	&	5837.0	&	5830.32(0.27)	&	5834	&	5835	&	5849	&	5830	\\
(2, 0, 1/2, 1)	&	\textit{$|2S, 1/2^+\rangle$}	&	6229.3	&		&	6213	&	6397	&	6225	&	6275	\\
(2, 0, 3/2, 1)	&	\textit{$|2S, 3/2^+\rangle$}	&	6234.3	&		&	6226	&	6421	&	6246	&	6291	\\
(3, 0, 1/2, 1)	&	\textit{$|3S, 1/2^+\rangle$}	&	6557.1	&		&	6575	&		&	6430	&	6707	\\
(3, 0, 3/2, 1)	&	\textit{$|3S, 3/2^+\rangle$}	&	6558.3	&		&	6583	&		&	6450	&	6720	\\
(4, 0, 1/2, 1)	&	\textit{$|4S, 1/2^+\rangle$}	&	6838.2	&		&	6869	&		&	6566	&	7113	\\
(4, 0, 3/2, 1)	&	\textit{$|4S, 3/2^+\rangle$}	&	6838.5	&		&	6876	&		&	6579	&	7124	\\
(5, 0, 1/2, 1)	&	\textit{$|5S, 1/2^+\rangle$}	&	7088.6	&		&	7124	&		&		&	7497	\\
(5, 0, 3/2, 1)	&	\textit{$|5S, 3/2^+\rangle$}	&	7088.7	&		&	7129	&		&		&	7506	\\
(6, 0, 1/2, 1)	&	\textit{$|6S, 1/2^+\rangle$}	&	7316.8	&		&		&		&		&	7862	\\
(6, 0, 3/2, 1)	&	\textit{$|6S, 3/2^+\rangle$}	&	7316.8	&		&		&		&		&	7869	\\
(1, 1, 1/2, 0)	&	\textit{$|1P, 1/2^-\rangle$}	&	6030.2	&		&	6095	&	6098	&	6113	&	6095	\\
(1, 1, 1/2, 1)	&	\textit{$|1P, 1/2^-\rangle$}	&	6075.0	&		&	6101	&	6113	&	6107	&	6101	\\
(1, 1, 3/2, 1)	&	\textit{$|1P, 3/2^-\rangle$}	&	6097.4	&	6095.80(1.7)	&	6087	&	6107	&	6116	&	6087	\\
(1, 1, 3/2, 2)	&	\textit{$|1P, 3/2^-\rangle$}	&	6201.3	&		&	6096	&	6122	&	6104	&	6105	\\
(1, 1, 5/2, 2)	&	\textit{$|1P, 5/2^-\rangle$}	&	6214.6	&		&	6084	&	6137	&	6119	&	6118	\\
(2, 1, 1/2, 0)	&	\textit{$|2P, 1/2^-\rangle$}	&	6434.4	&		&	6430	&		&	6447	&		\\
(2, 1, 1/2, 1)	&	\textit{$|2P, 1/2^-\rangle$}	&	6457.1	&		&	6440	&		&	6442	&		\\
(2, 1, 3/2, 1)	&	\textit{$|2P, 3/2^-\rangle$}	&	6463.6	&		&	6423	&		&	6450	&	6506	\\
(2, 1, 3/2, 2)	&	\textit{$|2P, 3/2^-\rangle$}	&	6513.2	&		&	6430	&		&	6439	&		\\
(2, 1, 5/2, 2)	&	\textit{$|2P, 5/2^-\rangle$}	&	6517.1	&		&	6421	&		&	6452	&	6489	\\
(3, 1, 1/2, 0)	&	\textit{$|3P, 1/2^-\rangle$}	&	6739.8	&		&	6742	&		&	6648	&		\\
(3, 1, 1/2, 1)	&	\textit{$|3P, 1/2^-\rangle$}	&	6756.5	&		&	6756	&		&	6643	&		\\
(3, 1, 3/2, 1)	&	\textit{$|3P, 3/2^-\rangle$}	&	6759.7	&		&	6736	&		&	6650	&	6884	\\
(3, 1, 3/2, 2)	&	\textit{$|3P, 3/2^-\rangle$}	&	6795.4	&		&	6742	&		&	6641	&		\\
(3, 1, 5/2, 2)	&	\textit{$|3P, 5/2^-\rangle$}	&	6797.1	&		&	6732	&		&	6652	&	6840	\\
(4, 1, 1/2, 0)	&	\textit{$|4P, 1/2^-\rangle$}	&	7003.7	&		&	7008	&		&	6739	&		\\
(4, 1, 1/2, 1)	&	\textit{$|4P, 1/2^-\rangle$}	&	7017.4	&		&	7024	&		&	6736	&		\\
(4, 1, 3/2, 1)	&	\textit{$|4P, 3/2^-\rangle$}	&	7019.4	&		&	7003	&		&	6741	&	7242	\\
(4, 1, 3/2, 2)	&	\textit{$|4P, 3/2^-\rangle$}	&	7048.3	&		&	7009	&		&	6734	&		\\
(4, 1, 5/2, 2)	&	\textit{$|4P, 5/2^-\rangle$}	&	7049.4	&		&	6999	&		&	6743	&	7174	\\
(1, 2, 1/2, 1)	&	\textit{$|1D, 1/2^+\rangle$}	&	6317.7	&		&	6311	&	6393	&	6338	&		\\
(1, 2, 3/2, 1)	&	\textit{$|1D, 3/2^+\rangle$}	&	6328.7	&		&	6285	&	6388	&	6344	&	6293	\\
(1, 2, 3/2, 2)	&	\textit{$|1D, 3/2^+\rangle$}	&	6379.9	&		&	6326	&	6403	&	6338	&	6375	\\
(1, 2, 5/2, 2)	&	\textit{$|1D, 5/2^+\rangle$}	&	6472.9	&		&	6270	&	6418	&	6345	&	6386	\\
(1, 2, 5/2, 3)	&	\textit{$|1D, 5/2^+\rangle$}	&	6390.2	&		&	6284	&	6404	&	6338	&	6346	\\
(1, 2, 7/2, 3)	&	\textit{$|1D, 7/2^+\rangle$}	&	6481.0	&		&	6260	&	6440	&	6346	&	6393	\\
(2, 2, 1/2, 1)	&	\textit{$|2D, 1/2^+\rangle$}	&	6650.6	&		&	6636	&		&	6639	&		\\
(2, 2, 3/2, 1)	&	\textit{$|2D, 3/2^+\rangle$}	&	6656.4	&		&	6612	&		&	6645	&		\\
(2, 2, 3/2, 2)	&	\textit{$|2D, 3/2^+\rangle$}	&	6691.8	&		&	6647	&		&	6639	&		\\
(2, 2, 5/2, 2)	&	\textit{$|2D, 5/2^+\rangle$}	&	6753.2	&		&	6598	&		&	6639	&	6778	\\
(2, 2, 5/2, 3)	&	\textit{$|2D, 5/2^+\rangle$}	&	6696.9	&		&	6612	&		&	6646	&		\\
(2, 2, 7/2, 3)	&	\textit{$|2D, 7/2^+\rangle$}	&	6757.0	&		&	6590	&		&	6647	&	6751	\\
(3, 2, 1/2, 1)	&	\textit{$|3D, 1/2^+\rangle$}	&	6927.9	&		&		&		&	6828	&		\\
(3, 2, 3/2, 1)	&	\textit{$|3D, 3/2^+\rangle$}	&	6931.7	&		&		&		&	6833	&		\\
(3, 2, 3/2, 2)	&	\textit{$|3D, 3/2^+\rangle$}	&	6960.1	&		&		&		&	6828	&		\\
(3, 2, 5/2, 2)	&	\textit{$|3D, 5/2^+\rangle$}	&	7008.0	&		&		&		&	6827	&	7148	\\
(3, 2, 5/2, 3)	&	\textit{$|3D, 5/2^+\rangle$}	&	6963.3	&		&		&		&	6833	&		\\
(3, 2, 7/2, 3)	&	\textit{$|3D, 7/2^+\rangle$}	&	7010.3	&		&		&		&	6834	&	7091	\\
(1, 3, 3/2, 2)	&	\textit{$|1F, 3/2^-\rangle$}	&	6558.7	&		&	6550	&		&		&		\\
(1, 3, 5/2, 2)	&	\textit{$|1F, 5/2^-\rangle$}	&	6567.2	&		&	6501	&		&		&		\\
(1, 3, 5/2, 3)	&	\textit{$|1F, 5/2^-\rangle$}	&	6624.6	&		&	6564	&		&		&		\\
(1, 3, 7/2, 3)	&	\textit{$|1F, 7/2^-\rangle$}	&	6712.3	&		&	6472	&		&		&	6655	\\
(1, 3, 7/2, 4)	&	\textit{$|1F, 7/2^-\rangle$}	&	6632.2	&		&	6500	&		&		&		\\
(1, 3, 9/2, 4)	&	\textit{$|1F, 9/2^-\rangle$}	&	6718.6	&		&	6459	&		&		&	6657	\\
(1, 4, 5/2, 3)	&	\textit{$|1G, 5/2^-\rangle$}	&	6770.4	&		&	6749	&		&		&		\\
(1, 4, 7/2, 3)	&	\textit{$|1G, 7/2^-\rangle$}	&	6777.7	&		&	6688	&		&		&		\\
(1, 4, 7/2, 4)	&	\textit{$|1G, 7/2^-\rangle$}	&	6840.6	&		&	6761	&		&		&		\\
(1, 4, 9/2, 4)	&	\textit{$|1G, 9/2^-\rangle$}	&	6928.3	&		&	6648	&		&		&	6913	\\
(1, 4, 9/2, 5)	&	\textit{$|1G, 9/2^-\rangle$}	&	6847.1	&		&	6687	&		&		&		\\
(1, 4, 11/2, 5)	&	\hspace{1.55mm}\textit{$|1G, 11/2^-\rangle$}	&	6933.9	&		&	6635	&		&		&	6910	\\

\end{tabular}
\end{ruledtabular}
\end{table*}  

\begin{table*}
\caption{\label{tab:table5}
	The same as Table \ref{tab:table2}, but with regard to the $\Xi_{b}^{'}$ baryonic states.}

\begin{ruledtabular}
\begin{tabular}{cccccccccccc}
$(n, L, J, j)$	&	States\textit{$|nL, J^P\rangle$}	&	Present	&	PDG \cite{Workman:2022ynf}	&	\cite{Ebert:2011kk}	&	\cite{Garcia-Tecocoatzi:2023btk}	&	\cite{li2022systematic}	&	\cite{Oudichhya:2021yln}	\\
\hline										   					
(1, 0, 1/2, 1)	&	\textit{$|1S, 1/2^+\rangle$}	&	5945.1	&	5935.02(0.05)	&	5936	&	5927	&	5943	&	5935	\\
(1, 0, 3/2, 1)	&	\textit{$|1S, 3/2^+\rangle$}	&	5962.7	&	5955.33(0.13)	&	5963	&	5951	&	5971	&	5952	\\
(2, 0, 1/2, 1)	&	\textit{$|2S, 1/2^+\rangle$}	&	6366.5	&		&	6329	&	6483	&	6350	&	6329	\\
(2, 0, 3/2, 1)	&	\textit{$|2S, 3/2^+\rangle$}	&	6371.4	&		&	6342	&	6507	&	6370	&	6316	\\
(3, 0, 1/2, 1)	&	\textit{$|3S, 1/2^+\rangle$}	&	6705.8	&		&	6687	&		&	6535	&	6700	\\
(3, 0, 3/2, 1)	&	\textit{$|3S, 3/2^+\rangle$}	&	6707.2	&		&	6695	&		&	6554	&	6660	\\
(4, 0, 1/2, 1)	&	\textit{$|4S, 1/2^+\rangle$}	&	6998.5	&		&	6978	&		&	6691	&	7051	\\
(4, 0, 3/2, 1)	&	\textit{$|4S, 3/2^+\rangle$}	&	6998.8	&		&	6984	&		&	6705	&	6987	\\
(5, 0, 1/2, 1)	&	\textit{$|5S, 1/2^+\rangle$}	&	7259.9	&		&	7229	&		&		&	7386	\\
(5, 0, 3/2, 1)	&	\textit{$|5S, 3/2^+\rangle$}	&	7260.0	&		&	7234	&		&		&	7300	\\
(6, 0, 1/2, 1)	&	\textit{$|6S, 1/2^+\rangle$}	&	7498.5	&		&		&		&		&	7706	\\
(6, 0, 3/2, 1)	&	\textit{$|6S, 3/2^+\rangle$}	&	7498.5	&		&		&		&		&	7600	\\
(1, 1, 1/2, 0)	&	\textit{$|1P, 1/2^-\rangle$}	&	6185.1	&		&	6227	&	6199	&	6238	&	6227	\\
(1, 1, 1/2, 1)	&	\textit{$|1P, 1/2^-\rangle$}	&	6219.7	&	6227.9(0.9)	&	6233	&	6213	&	6232	&	6233	\\
(1, 1, 3/2, 1)	&	\textit{$|1P, 3/2^-\rangle$}	&	6242.0	&	6227.9(0.9)	&	6224	&	6208	&	6240	&	6224	\\
(1, 1, 3/2, 2)	&	\textit{$|1P, 3/2^-\rangle$}	&	6325.7	&	6327.28(0.35)	&	6234	&	6223	&	6229	&	6229	\\
(1, 1, 5/2, 2)	&	\textit{$|1P, 5/2^-\rangle$}	&	6338.9	&	6332.69(0.28)	&	6226	&	6238	&	6243	&	6240	\\
(2, 1, 1/2, 0)	&	\textit{$|2P, 1/2^-\rangle$}	&	6591.1	&		&	6604	&		&	6569	&		\\
(2, 1, 1/2, 1)	&	\textit{$|2P, 1/2^-\rangle$}	&	6608.5	&		&	6611	&		&	6564	&		\\
(2, 1, 3/2, 1)	&	\textit{$|2P, 3/2^-\rangle$}	&	6615.1	&		&	6598	&		&	6572	&	6605	\\
(2, 1, 3/2, 2)	&	\textit{$|2P, 3/2^-\rangle$}	&	6654.0	&		&	6605	&		&	6562	&		\\
(2, 1, 5/2, 2)	&	\textit{$|2P, 5/2^-\rangle$}	&	6658.1	&		&	6596	&		&	6574	&	6451	\\
(3, 1, 1/2, 0)	&	\textit{$|3P, 1/2^-\rangle$}	&	6905.3	&		&	6906	&		&	6758	&		\\
(3, 1, 1/2, 1)	&	\textit{$|3P, 1/2^-\rangle$}	&	6918.1	&		&	6915	&		&	6754	&		\\
(3, 1, 3/2, 1)	&	\textit{$|3P, 3/2^-\rangle$}	&	6921.4	&		&	6900	&		&	6760	&	6961	\\
(3, 1, 3/2, 2)	&	\textit{$|3P, 3/2^-\rangle$}	&	6949.1	&		&	6905	&		&	6752	&		\\
(3, 1, 5/2, 2)	&	\textit{$|3P, 5/2^-\rangle$}	&	6951.0	&		&	6897	&		&	6762	&	6655	\\
(4, 1, 1/2, 0)	&	\textit{$|4P, 1/2^-\rangle$}	&	7178.9	&		&	7164	&		&	6866	&		\\
(4, 1, 1/2, 1)	&	\textit{$|4P, 1/2^-\rangle$}	&	7189.4	&		&	7174	&		&	6863	&		\\
(4, 1, 3/2, 1)	&	\textit{$|4P, 3/2^-\rangle$}	&	7191.5	&		&	7159	&		&	6868	&	7299	\\
(4, 1, 3/2, 2)	&	\textit{$|4P, 3/2^-\rangle$}	&	7213.8	&		&	7163	&		&	6861	&		\\
(4, 1, 5/2, 2)	&	\textit{$|4P, 5/2^-\rangle$}	&	7215.0	&		&	7156	&		&	6869	&	6853	\\
(1, 2, 1/2, 1)	&	\textit{$|1D, 1/2^+\rangle$}	&	6478.7	&		&	6447	&	6479	&	6460	&		\\
(1, 2, 3/2, 1)	&	\textit{$|1D, 3/2^+\rangle$}	&	6489.7	&		&	6431	&	6474	&	6466	&	6425	\\
(1, 2, 3/2, 2)	&	\textit{$|1D, 3/2^+\rangle$}	&	6529.1	&		&	6459	&	6488	&	6460	&	6508	\\
(1, 2, 5/2, 2)	&	\textit{$|1D, 5/2^+\rangle$}	&	6604.6	&		&	6420	&	6489	&	6466	&	6484	\\
(1, 2, 5/2, 3)	&	\textit{$|1D, 5/2^+\rangle$}	&	6539.6	&		&	6432	&	6504	&	6460	&	6510	\\
(1, 2, 7/2, 3)	&	\textit{$|1D, 7/2^+\rangle$}	&	6612.9	&		&	6414	&	6526	&	6467	&	6516	\\
(2, 2, 1/2, 1)	&	\textit{$|2D, 1/2^+\rangle$}	&	6818.1	&		&	6767	&		&	6757	&		\\
(2, 2, 3/2, 1)	&	\textit{$|2D, 3/2^+\rangle$}	&	6824.0	&		&	6751	&		&	6763	&		\\
(2, 2, 3/2, 2)	&	\textit{$|2D, 3/2^+\rangle$}	&	6851.0	&		&	6775	&		&	6758	&		\\
(2, 2, 5/2, 2)	&	\textit{$|2D, 5/2^+\rangle$}	&	6899.9	&		&	6751	&		&	6764	&	6751	\\
(2, 2, 5/2, 3)	&	\textit{$|2D, 5/2^+\rangle$}	&	6856.2	&		&	6740	&		&	6757	&		\\
(2, 2, 7/2, 3)	&	\textit{$|2D, 7/2^+\rangle$}	&	6903.9	&		&	6736	&		&	6765	&	6672	\\
(3, 2, 1/2, 1)	&	\textit{$|3D, 1/2^+\rangle$}	&	7104.3	&		&		&		&	6941	&		\\
(3, 2, 3/2, 1)	&	\textit{$|3D, 3/2^+\rangle$}	&	7108.1	&		&		&		&	6946	&		\\
(3, 2, 3/2, 2)	&	\textit{$|3D, 3/2^+\rangle$}	&	7129.7	&		&		&		&	6941	&		\\
(3, 2, 5/2, 2)	&	\textit{$|3D, 5/2^+\rangle$}	&	7167.5	&		&		&		&	6946	&	6984	\\
(3, 2, 5/2, 3)	&	\textit{$|3D, 5/2^+\rangle$}	&	7133.0	&		&		&		&	6941	&		\\
(3, 2, 7/2, 3)	&	\textit{$|3D, 7/2^+\rangle$}	&	7170.0	&		&		&		&	6946	&	6824	\\
(1, 3, 3/2, 2)	&	\textit{$|1F, 3/2^-\rangle$}	&	6728.8	&		&	6675	&		&	6657	&		\\
(1, 3, 5/2, 2)	&	\textit{$|1F, 5/2^-\rangle$}	&	6737.3	&		&	6640	&		&	6660	&	6612	\\
(1, 3, 5/2, 3)	&	\textit{$|1F, 5/2^-\rangle$}	&	6781.4	&		&	6686	&		&	6657	&	6777	\\
(1, 3, 7/2, 3)	&	\textit{$|1F, 7/2^-\rangle$}	&	6851.4	&		&	6641	&		&	6660	&	6779	\\
(1, 3, 7/2, 4)	&	\textit{$|1F, 7/2^-\rangle$}	&	6789.2	&		&	6619	&		&	6657	&	6734	\\
(1, 3, 9/2, 4)	&	\textit{$|1F, 9/2^-\rangle$}	&	6858.0	&		&	6610	&		&	6661	&	6780	\\
(1, 4, 5/2, 3)	&	\textit{$|1G, 5/2^-\rangle$}	&	6949.9	&		&	6867	&		&		&		\\
(1, 4, 7/2, 3)	&	\textit{$|1G, 7/2^-\rangle$}	&	6957.2	&		&	6822	&		&		&	6794	\\
(1, 4, 7/2, 4)	&	\textit{$|1G, 7/2^-\rangle$}	&	7005.5	&		&	6876	&		&		&	7036	\\
(1, 4, 9/2, 4)	&	\textit{$|1G, 9/2^-\rangle$}	&	7074.9	&		&	6821	&		&		&	7038	\\
(1, 4, 9/2, 5)	&	\textit{$|1G, 9/2^-\rangle$}	&	7012.1	&		&	6792	&		&		&	6974	\\
(1, 4, 11/2, 5)	&	\hspace{1.55mm}\textit{$|1G, 11/2^-\rangle$}	&	7080.7	&		&	6782	&		&		&	7034	\\

\end{tabular}
\end{ruledtabular}
\end{table*}  

\begin{table*}
\caption{\label{tab:table6}
The same as Table \ref{tab:table2}, but with regard to the $\Omega_b$ baryonic states.}

\begin{ruledtabular}
\begin{tabular}{ccccccccc}
$(n, L, J, j)$  &       States\textit{$|nL, J^P\rangle$}       & Present & PDG \cite{Workman:2022ynf} & \cite{Ebert:2011kk} & \cite{Garcia-Tecocoatzi:2023btk} & \cite{YU2023116183} & \cite{Oudichhya:2021yln} &  \\ \hline
(1, 0, 1/2, 1)  &         \textit{$|1S, 1/2^+\rangle$}         & 6065.2  &        6045.2(1.2)        &        6064         &               6059               &        6043         &           6054           &  \\
(1, 0, 3/2, 1)  &         \textit{$|1S, 3/2^+\rangle$}         & 6080.6  &                           &        6088         &               6083               &        6069         &           6074           &  \\
(2, 0, 1/2, 1)  &         \textit{$|2S, 1/2^+\rangle$}         & 6492.0  &                           &        6450         &               6590               &        6446         &           6455           &  \\
(2, 0, 3/2, 1)  &         \textit{$|2S, 3/2^+\rangle$}         & 6496.8  &                           &        6461         &               6614               &        6466         &           6481           &  \\
(3, 0, 1/2, 1)  &         \textit{$|3S, 1/2^+\rangle$}         & 6839.9  &                           &        6804         &                                  &        6633         &           6832           &  \\
(3, 0, 3/2, 1)  &         \textit{$|3S, 3/2^+\rangle$}         & 6841.4  &                           &        6811         &                                  &        6650         &           6864           &  \\
(4, 0, 1/2, 1)  &         \textit{$|4S, 1/2^+\rangle$}         & 7141.4  &                           &        7091         &                                  &        6790         &           7190           &  \\
(4, 0, 3/2, 1)  &         \textit{$|4S, 3/2^+\rangle$}         & 7141.8  &                           &        7096         &                                  &        6804         &           7226           &  \\
(5, 0, 1/2, 1)  &         \textit{$|5S, 1/2^+\rangle$}         & 7411.5  &                           &        7338         &                                  &                     &           7531           &  \\
(5, 0, 3/2, 1)  &         \textit{$|5S, 3/2^+\rangle$}         & 7411.6  &                           &        7343         &                                  &                     &           7572           &  \\
(6, 0, 1/2, 1)  &         \textit{$|6S, 1/2^+\rangle$}         & 7658.5  &                           &                     &                                  &                     &           7857           &  \\
(6, 0, 3/2, 1)  &         \textit{$|6S, 3/2^+\rangle$}         & 7658.5  &                           &                     &                                  &                     &           7902           &  \\
(1, 1, 1/2, 0)  &         \textit{$|1P, 1/2^-\rangle$}         & 6322.0  &   6315.6(0.6)                   &        6330         &               6318               &        6334         &           6359           &  \\
(1, 1, 1/2, 1)  &         \textit{$|1P, 1/2^-\rangle$}         & 6350.1  &   6330.3(0.6)                        &        6339         &               6333               &        6329         &           6365           &  \\
(1, 1, 3/2, 1)  &         \textit{$|1P, 3/2^-\rangle$}         & 6372.3  &   6339.7(0.6)                        &        6331         &               6328               &        6336         &           6348           &  \\
(1, 1, 3/2, 2)  &         \textit{$|1P, 3/2^-\rangle$}         & 6442.7  & 6349.8(0.6)                          &        6340         &               6342               &        6326         &           6360           &  \\
(1, 1, 5/2, 2)  &         \textit{$|1P, 5/2^-\rangle$}         & 6455.8  &                           &        6334         &               6358               &        6339         &           6362           &  \\
(2, 1, 1/2, 0)  &         \textit{$|2P, 1/2^-\rangle$}         & 6730.0  &                           &        6706         &                                  &        6662         &                          &  \\
(2, 1, 1/2, 1)  &         \textit{$|2P, 1/2^-\rangle$}         & 6743.9  &                           &        6710         &                                  &        6658         &                          &  \\
(2, 1, 3/2, 1)  &         \textit{$|2P, 3/2^-\rangle$}         & 6750.6  &                           &        6699         &                                  &        6664         &           6662           &  \\
(2, 1, 3/2, 2)  &         \textit{$|2P, 3/2^-\rangle$}         & 6782.6  &                           &        6705         &                                  &        6655         &                          &  \\
(2, 1, 5/2, 2)  &         \textit{$|2P, 5/2^-\rangle$}         & 6786.8  &                           &        6700         &                                  &        6666         &           6653           &  \\
(3, 1, 1/2, 0)  &         \textit{$|3P, 1/2^-\rangle$}         & 7051.3  &                           &        7003         &                                  &        6844         &                          &  \\
(3, 1, 1/2, 1)  &         \textit{$|3P, 1/2^-\rangle$}         & 7061.5  &                           &        7009         &                                  &        6841         &                          &  \\
(3, 1, 3/2, 1)  &         \textit{$|3P, 3/2^-\rangle$}         & 7064.9  &                           &        6998         &                                  &        6846         &           6962           &  \\
(3, 1, 3/2, 2)  &         \textit{$|3P, 3/2^-\rangle$}         & 7087.5  &                           &        7002         &                                  &        6839         &                          &  \\
(3, 1, 5/2, 2)  &         \textit{$|3P, 5/2^-\rangle$}         & 7089.5  &                           &        6996         &                                  &        6848         &           6932           &  \\
(4, 1, 1/2, 0)  &         \textit{$|4P, 1/2^-\rangle$}         & 7332.7  &                           &        7257         &                                  &        6969         &                          &  \\
(4, 1, 1/2, 1)  &         \textit{$|4P, 1/2^-\rangle$}         & 7341.2  &                           &        7265         &                                  &        6966         &                          &  \\
(4, 1, 3/2, 1)  &         \textit{$|4P, 3/2^-\rangle$}         & 7343.2  &                           &        7250         &                                  &        6970         &           7249           &  \\
(4, 1, 3/2, 2)  &         \textit{$|4P, 3/2^-\rangle$}         & 7361.4  &                           &        7258         &                                  &        6964         &                          &  \\
(4, 1, 5/2, 2)  &         \textit{$|4P, 5/2^-\rangle$}         & 7362.6  &                           &        7251         &                                  &        6972         &           7200           &  \\
(1, 2, 1/2, 1)  &         \textit{$|1D, 1/2^+\rangle$}         & 6620.1  &                           &        6540         &               6585               &        6556         &                          &  \\
(1, 2, 3/2, 1)  &         \textit{$|1D, 3/2^+\rangle$}         & 6631.2  &                           &        6530         &               6581               &        6561         &           6557           &  \\
(1, 2, 3/2, 2)  &         \textit{$|1D, 3/2^+\rangle$}         & 6662.9  &                           &        6549         &               6595               &        6556         &           6640           &  \\
(1, 2, 5/2, 2)  &         \textit{$|1D, 5/2^+\rangle$}         & 6727.0  &                           &        6529         &               6610               &        6561         &           6629           &  \\
(1, 2, 5/2, 3)  &         \textit{$|1D, 5/2^+\rangle$}         & 6673.5  &                           &        6520         &               6596               &        6555         &           6620           &  \\
(1, 2, 7/2, 3)  &         \textit{$|1D, 7/2^+\rangle$}         & 6735.6  &                           &        6517         &               6632               &        6562         &           6638           &  \\
(2, 2, 1/2, 1)  &         \textit{$|2D, 1/2^+\rangle$}         & 6965.0  &                           &        6857         &                                  &        6846         &                          &  \\
(2, 2, 3/2, 1)  &         \textit{$|2D, 3/2^+\rangle$}         & 6970.9  &                           &        6846         &                                  &        6852         &                          &  \\
(2, 2, 3/2, 2)  &         \textit{$|2D, 3/2^+\rangle$}         & 6992.5  &                           &        6863         &                                  &        6846         &                          &  \\
(2, 2, 5/2, 2)  &         \textit{$|2D, 5/2^+\rangle$}         & 7033.5  &                           &        6846         &                                  &        6852         &           6659           &  \\
(2, 2, 5/2, 3)  &         \textit{$|2D, 5/2^+\rangle$}         & 6997.8  &                           &        6837         &                                  &        6846         &                          &  \\
(2, 2, 7/2, 3)  &         \textit{$|2D, 7/2^+\rangle$}         & 7037.6  &                           &        6834         &                                  &        6853         &           6643           &  \\
(3, 2, 1/2, 1)  &         \textit{$|3D, 1/2^+\rangle$}         & 7258.4  &                           &                     &                                  &        7021         &                          &  \\
(3, 2, 3/2, 1)  &         \textit{$|3D, 3/2^+\rangle$}         & 7262.3  &                           &                     &                                  &        7026         &                          &  \\
(3, 2, 3/2, 2)  &         \textit{$|3D, 3/2^+\rangle$}         & 7279.6  &                           &                     &                                  &        7022         &                          &  \\
(3, 2, 5/2, 2)  &         \textit{$|3D, 5/2^+\rangle$}         & 7310.9  &                           &                     &                                  &        7026         &           6689           &  \\
(3, 2, 5/2, 3)  &         \textit{$|3D, 5/2^+\rangle$}         & 7282.9  &                           &                     &                                  &        7021         &                          &  \\
(3, 2, 7/2, 3)  &         \textit{$|3D, 7/2^+\rangle$}         & 7313.5  &                           &                     &                                  &        7027         &           6648           &  \\
(1, 3, 3/2, 2)  &         \textit{$|1F, 3/2^-\rangle$}         & 6877.0  &                           &        6763         &                                  &        6751         &                          &  \\
(1, 3, 5/2, 2)  &         \textit{$|1F, 5/2^-\rangle$}         & 6885.5  &                           &        6737         &                                  &        6754         &           6744           &  \\
(1, 3, 5/2, 3)  &         \textit{$|1F, 5/2^-\rangle$}         & 6921.1  &                           &        6771         &                                  &        6751         &           6909           &  \\
(1, 3, 7/2, 3)  &         \textit{$|1F, 7/2^-\rangle$}         & 6979.8  &                           &        6736         &                                  &        6754         &           6899           &  \\
(1, 3, 7/2, 4)  &         \textit{$|1F, 7/2^-\rangle$}         & 6929.0  &                           &        6719         &                                  &        6750         &           6870           &  \\
(1, 3, 9/2, 4)  &         \textit{$|1F, 9/2^-\rangle$}         & 6986.5  &                           &        6713         &                                  &        6754         &           6903           &  \\
(1, 4, 5/2, 3)  &         \textit{$|1G, 5/2^-\rangle$}         & 7105.3  &                           &        6952         &                                  &        6923         &                          &  \\
(1, 4, 7/2, 3)  &         \textit{$|1G, 7/2^-\rangle$}         & 7112.7  &                           &        6916         &                                  &        6925         &           6926           &  \\
(1, 4, 7/2, 4)  &         \textit{$|1G, 7/2^-\rangle$}         & 7151.5  &                           &        6959         &                                  &        6923         &           7168           &  \\
(1, 4, 9/2, 4)  &         \textit{$|1G, 9/2^-\rangle$}         & 7209.2  &                           &        6915         &                                  &        6925         &           7159           &  \\
(1, 4, 9/2, 5)  &         \textit{$|1G, 9/2^-\rangle$}         & 7158.3  &                           &        6892         &                                  &        6922         &           7111           &  \\
(1, 4, 11/2, 5) & \hspace{1.55mm}\textit{$|1G, 11/2^-\rangle$} & 7215.2  &                           &        6884         &                                  &        6925         &           7158           &
\end{tabular}
\end{ruledtabular}
\end{table*}
\newpage

\section{STRONG DECAY}

Heavy hadron chiral perturbation theory (HHChPT), which incorporates heavy-quark symmetry and chiral symmetry, provides the most convenient description for the strong decays of singly bottom baryons into another singly bottom baryon and a light pseudoscalar meson. In this approach, the strong decay width expressions for the $1S$- and $1P$- wave states of singly bottom baryons, derived from the relevant  chiral Lagrangian, are presented as follows\cite{PhysRevD.75.014006,PhysRevD.95.094018}:

\begin{widetext}
\begin{equation}
\label{deq:1}
\begin{split}
\Gamma[\Xi_{b}^{-}{|1P,1/2^-\rangle}]&=\Gamma[\Xi_{b}^{-}{|1P,1/2^-\rangle}\rightarrow\Xi_{b}^{'-}\pi^{0},\Xi_{b}^{'0}\pi^{-}]\\
&=\frac{h_{2}^{2}}{2\pi f_{\pi}^{2}} \left(\frac{1}{4}\frac{M_{\Xi_{b}^{'-}}}{M_{\Xi_{b}^{-}{|1P,1/2^-\rangle}}}E_{\pi^{0}}^{2}p_{\pi^{0}}+\frac{1}{2}\frac{M_{\Xi_{b}^{'0}}}{M_{\Xi_{b}^{-}{|1P,1/2^-\rangle}}}E_{\pi^{-}}^{2}p_{\pi^{-}}\right),
\end{split}
\end{equation}

\begin{equation}
\label{deq:2}
\begin{split}
	\Gamma[\Xi_{b}^{-}{|1P,3/2^-\rangle}]&=\Gamma[\Xi_{b}^{-}{|1P,3/2^-\rangle}\rightarrow\Xi_{b}^{'-}\pi^{0},\Xi_{b}^{'0}\pi^{-},\Xi_{b}^{*-}\pi^{0},\Xi_{b}^{*0}\pi^{-}]\\
	&=\frac{2 h_{8}^{2}}{9\pi f_{\pi}^{2}} \left(\frac{1}{4}\frac{M_{\Xi_{b}^{'-}}}{M_{\Xi_{b}^{-}{|1P,3/2^-\rangle}}}p_{\pi^{0}}^{5}+\frac{1}{2}\frac{M_{\Xi_{b}^{'0}}}{M_{\Xi_{b}^{-}{|1P,3/2^-\rangle}}}p_{\pi^{-}}^{5}\right)\\
	&+\frac{h_{2}^{2}}{2\pi f_{\pi}^{2}} \left(\frac{1}{4}\frac{M_{\Xi_{b}^{*-}}}{M_{\Xi_{b}^{-}{|1P,3/2^-\rangle}}}E_{\pi^{0}}^{2}p_{\pi^{0}}+\frac{1}{2}\frac{M_{\Xi_{b}^{*0}}}{M_{\Xi_{b}^{-}{|1P,3/2^-\rangle}}}E_{\pi^{-}}^{2}p_{\pi^{-}}\right),
\end{split}
\end{equation}

\begin{equation}
	\label{deq:3}
\begin{split}
\Gamma[\Sigma_{b}^{+}{|1S,1/2^+\rangle}]&=\Gamma[\Sigma_{b}^{+}{|1S,1/2^+\rangle}\rightarrow\Lambda_{b}^{0}\pi^{+}]\\
&=\frac{g_{2}^{2}}{2\pi f_{\pi}^{2}} \frac{M_{\Lambda_{b}^{0}}}{M_{\Sigma_{b}^{+}{|1S,1/2^+\rangle}}}p_{\pi^{+}}^{3},
\end{split}
\end{equation}

\begin{equation}
	\label{deq:4}
\begin{split}	
	\Gamma[\Sigma_{b}^{+}{|1S,3/2^+\rangle}]&=\Gamma[\Sigma_{b}^{+}{|1S,3/2^+\rangle}\rightarrow\Lambda_{b}^{0}\pi^{+}]\\
	&=\frac{g_{2}^{2}}{2\pi f_{\pi}^{2}} \frac{M_{\Lambda_{b}^{0}}}{M_{\Sigma_{b}^{+}{|1S,3/2^+\rangle}}}p_{\pi^{+}}^{3},
\end{split}
\end{equation}

\begin{equation}
	\label{deq:5}
\begin{split}
	\Gamma[\Sigma_{b}^{+}|1P,1/2^-\rangle_{j=0}]&=\Gamma[\Sigma_{b}^{+}|1P,1/2^-\rangle_{j=0}\rightarrow\Lambda_{b}^{0}\pi^{+}]\\
	&=\frac{h_{3}^{2}}{2\pi f_{\pi}^{2}} \frac{M_{\Lambda_{b}^{0}}}{M_{\Sigma_{b}^{+}|1P,1/2^-\rangle_{j=0}}}E_{\pi^{+}}^{2}p_{\pi^{+}},
\end{split}
\end{equation}

\begin{equation}
	\label{deq:6}
\begin{split}
	\Gamma[\Sigma_{b}^{+}|1P,1/2^{-}\rangle_{j=1}]&=\Gamma[\Sigma_{b}^{+}|1P,1/2^{-}\rangle_{j=1}\rightarrow\Sigma_{b}^{+}\pi^{0}, \Sigma_{b}^{0}\pi^{+}]\\
	&=\frac{h_{4}^{2}}{4\pi f_{\pi}^{2}}\left(\frac{M_{\Sigma_{b}^{+}}}{M_{\Sigma_{b}^{+}|1P,1/2^-\rangle_{j=1}}}E_{\pi^{0}}^{2}p_{\pi^{0}}+\frac{M_{\Sigma_{b}^{0}}}{M_{\Sigma_{b}^{+}|1P,1/2^-\rangle_{j=1}}}E_{\pi^{+}}^{2}p_{\pi^{+}}	\right),
\end{split}
\end{equation}

\begin{equation}
	\label{deq:7}
\begin{split}
	\Gamma[\Sigma_{b}^{+}|1P,3/2^-\rangle_{j=1}]&=\Gamma[\Sigma_{b}^{+}|1P,3/2^-\rangle_{j=1}\rightarrow\Sigma_{b}^{*+}\pi^{0}, \Sigma_{b}^{*0}\pi^{+}]\\
	&=\frac{h_{9}^{2}}{9\pi f_{\pi}^{2}}\left(\frac{M_{\Sigma_{b}^{*+}}}{M_{\Sigma_{b}^{+}|1P,3/2^-\rangle_{j=1}}}p_{\pi^{0}}^{5}+\frac{M_{\Sigma_{b}^{*0}}}{M_{\Sigma_{b}^{+}|1P,3/2^-\rangle_{j=1}}}p_{\pi^{+}}^{5}	\right),
\end{split}
\end{equation}

\begin{equation}
	\label{deq:8}
\begin{split}
	\Gamma[\Xi_{b}^{'-}|1S,3/2^+\rangle]&=\Gamma[\Xi_{b}^{'-}|1S,3/2^+\rangle\rightarrow\Xi_{b}^{-}\pi^{0},\Xi_{b}^{0}\pi^{-}]\\
	&=\frac{g_{2}^{2}}{2\pi f_{\pi}^{2}} \left(\frac{1}{4}\frac{M_{\Xi_{b}^{-}}}{M_{\Xi_{b}^{'-}|1S,3/2^+\rangle}}p_{\pi^{0}}^{3}+\frac{1}{2}\frac{M_{\Xi_{b}^{0}}}{M_{\Xi_{b}^{'-}|1S,3/2^+\rangle}}p_{\pi^{-}}^{3}\right),
\end{split}
\end{equation}

\begin{equation}
	\label{deq:9}
	\begin{split}
		\Gamma[\Xi_{b}^{'-}|1P,1/2^-\rangle_{j=0}]&=\Gamma[\Xi_{b}^{'-}|1P,1/2^-\rangle_{j=0}\rightarrow\Xi_{b}^{-}\pi^{0},\Xi_{b}^{0}\pi^{-}]\\
		&=\frac{h_{3}^{2}}{2\pi f_{\pi}^{2}} \left(\frac{1}{4}\frac{M_{\Xi_{b}^{-}}}{M_{\Xi_{b}^{'-}|1P,1/2^-\rangle_{j=0}}}E_{\pi^{0}}^{2}p_{\pi^{0}}+\frac{1}{2}\frac{M_{\Xi_{b}^{0}}}{M_{\Xi_{b}^{'-}|1P,1/2^-\rangle_{j=0}}}E_{\pi^{-}}^{2}p_{\pi^{-}}\right),
	\end{split}
\end{equation}

\begin{equation}
	\label{deq:10}
\begin{split}
	\Gamma[\Xi_{b}^{'-}|1P,1/2^-\rangle_{j=1}]&=\Gamma[\Xi_{b}^{'-}|1P,1/2^-\rangle_{j=1}\rightarrow\Xi_{b}^{'-}\pi^{0},\Xi_{b}^{'0}\pi^{-}]\\
	&=\frac{h_{4}^{2}}{4\pi f_{\pi}^{2}} \left(\frac{1}{4}\frac{M_{\Xi_{b}^{'-}}}{M_{\Xi_{b}^{'-}|1P,1/2^-\rangle_{j=1}}}E_{\pi^{0}}^{2}p_{\pi^{0}}+\frac{1}{2}\frac{M_{\Xi_{b}^{'0}}}{M_{\Xi_{b}^{'-}|1P,1/2^-\rangle_{j=1}}}E_{\pi^{-}}^{2}p_{\pi^{-}}\right),
\end{split}
\end{equation}

\begin{equation}
	\label{deq:11}
\begin{split}
	\Gamma[\Xi_{b}^{'-}|1P,3/2^-\rangle_{j=1}]&=\Gamma[\Xi_{b}^{'-}|1P,3/2^-\rangle_{j=1}\rightarrow\Xi_{b}^{'-}\pi^{0},\Xi_{b}^{'0}\pi^{-}]\\
	&=\frac{h_{9}^{2}}{9\pi f_{\pi}^{2}} \left(\frac{1}{4}\frac{M_{\Xi_{b}^{'-}}}{M_{\Xi_{b}^{'-}|1P,3/2^-\rangle_{j=1}}}p_{\pi^{0}}^{5}+\frac{1}{2}\frac{M_{\Xi_{b}^{'0}}}{M_{\Xi_{b}^{'-}|1P,3/2^-\rangle_{j=1}}}p_{\pi^{-}}^{5}\right),
\end{split}
\end{equation}

\begin{equation}
	\label{deq:12}
	\begin{split}
		\Gamma[\Omega_{b}^{-}|1P,1/2^-\rangle_{j=0}]&=\Gamma[[\Omega_{b}^{-}|1P,1/2^-\rangle_{j=0}\rightarrow\Xi_{b}^{-}K^{0},\Xi_{b}^{0}K^{-}]\\
		&=\frac{h_{3}^{2}}{2\pi f_{\pi}^{2}} \left(\frac{M_{\Xi_{b}^{-}}}{M_{\Omega_{b}^{-}|1P,1/2^-\rangle_{j=0}}}E_{K^{0}}^{2}p_{K^{0}}+\frac{M_{\Xi_{b}^{0}}}{M_{\Omega_{b}^{-}|1P,1/2^-\rangle_{j=0}}}E_{K^{-}}^{2}p_{K^{-}}\right),
	\end{split}
\end{equation}

\begin{equation}
	\label{deq:13}
	\begin{split}
		\Gamma[\Omega_{b}^{-}|1P,1/2^-\rangle_{j=1}]&=\Gamma[\Omega_{b}^{-}|1P,1/2^-\rangle_{j=1}\rightarrow\Xi_{b}^{'-}K^{0},\Xi_{b}^{'0}K^{-}]\\
		&=\frac{h_{4}^{2}}{4\pi f_{\pi}^{2}} \left(\frac{M_{\Xi_{b}^{'-}}}{M_{\Omega_{b}^{-}|1P,1/2^-\rangle_{j=1}}}E_{K^{0}}^{2}p_{K^{0}}+\frac{M_{\Xi_{b}^{'0}}}{M_{\Omega_{b}^{-}|1P,1/2^-\rangle_{j=1}}}E_{K^{-}}^{2}p_{K^{-}}\right),
	\end{split}
\end{equation}

\begin{equation}
	\label{deq:14}
	\begin{split}
		\Gamma[\Omega_{b}^{-}|1P,3/2^-\rangle_{j=1}]&=\Gamma[\Omega_{b}^{-}|1P,3/2^-\rangle_{j=1}\rightarrow\Xi_{b}^{'-}K^{0},\Xi_{b}^{'0}K^{-}]\\
		&=\frac{h_{9}^{2}}{9\pi f_{\pi}^{2}} \left(\frac{M_{\Xi_{b}^{'-}}}{M_{\Omega_{b}^{-}|1P,3/2^-\rangle_{j=1}}}p_{K^{0}}^{5}+\frac{M_{\Xi_{b}^{'0}}}{M_{\Omega_{b}^{-}|1P,3/2^-\rangle_{j=1}}}p_{K^{-}}^{5}\right),
	\end{split}
\end{equation}

\begin{equation}
	\label{deq:15}
	\begin{split}
		\Gamma[\Omega_{b}^{-}|1P,3/2^-\rangle_{j=2}]&=\Gamma[\Omega_{b}^{-}|1P,3/2^-\rangle_{j=2}\rightarrow\Xi_{b}^{-}K^{0},\Xi_{b}^{0}K^{-},\Xi_{b}^{'-}K^{0},\Xi_{b}^{'0}K^{-}]\\
		&=\frac{4h_{10}^{2}}{15\pi f_{\pi}^{2}} \left(\frac{M_{\Xi_{b}^{-}}}{M_{\Omega_{b}^{-}|1P,3/2^-\rangle_{j=2}}}p_{K^{0}}^{5}+\frac{M_{\Xi_{b}^{0}}}{M_{\Omega_{b}^{-}|1P,3/2^-\rangle_{j=2}}}p_{K^{-}}^{5}\right)\\
		&+\frac{h_{11}^{2}}{10\pi f_{\pi}^{2}} \left(\frac{M_{\Xi_{b}^{'-}}}{M_{\Omega_{b}^{-}|1P,3/2^-\rangle_{j=2}}}p_{K^{0}}^{5}+\frac{M_{\Xi_{b}^{'0}}}{M_{\Omega_{b}^{-}|1P,3/2^-\rangle_{j=2}}}p_{K^{-}}^{5}\right),
	\end{split}
\end{equation}

\begin{equation}
	\label{deq:16}
	\begin{split}
		\Gamma[\Omega_{b}^{-}|1P,5/2^-\rangle_{j=2}]&=\Gamma[\Omega_{b}^{-}|1P,5/2^-\rangle_{j=2}\rightarrow\Xi_{b}^{-}K^{0},\Xi_{b}^{0}K^{-},\Xi_{b}^{'-}K^{0},\Xi_{b}^{'0}K^{-}]\\
		&=\frac{4h_{10}^{2}}{15\pi f_{\pi}^{2}} \left(\frac{M_{\Xi_{b}^{-}}}{M_{\Omega_{b}^{-}|1P,5/2^-\rangle_{j=2}}}p_{K^{0}}^{5}+\frac{M_{\Xi_{b}^{0}}}{M_{\Omega_{b}^{-}|1P,5/2^-\rangle_{j=2}}}p_{K^{-}}^{5}\right)\\
		&+\frac{2h_{11}^{2}}{45\pi f_{\pi}^{2}} \left(\frac{M_{\Xi_{b}^{'-}}}{M_{\Omega_{b}^{-}|1P,5/2^-\rangle_{j=2}}}p_{K^{0}}^{5}+\frac{M_{\Xi_{b}^{'0}}}{M_{\Omega_{b}^{-}|1P,5/2^-\rangle_{j=2}}}p_{K^{-}}^{5}\right).
	\end{split}
\end{equation}

\end{widetext}
where $p_{\pi/K}$ and $E_{\pi/K}$ represent the pion or kaon's center of mass momentum and energy, respectively. $f_{\pi}$=132. $g_{2}$ is the coupling constant for  the $P$-wave transition, $h_{2}-h_{4}$ are the coupling constants for the $S$-wave transition, and $h_{8}-h_{11}$ are the coupling constants for the $D$-wave transition. In the framework of HHChPT,  $g_{2}=0.591$, $h_{2}=0.437$, and $h_{8}<0.0365$MeV$^{-1}$ \cite{PhysRevD.75.014006}. According to the quark model, other coupling constants are related to $h_{2}$ or $h_{8}$ by \cite{PhysRevD.56.5483} 
\begin{equation}
	\begin{split}
	&|h_{3}|=\sqrt{3}|h_{2}|, \hspace{1.189cm} |h_{4}|=2|h_{2}|, \\
	&|h_{8}|=|h_{9}|=|h_{10}|, \hspace{0.5cm}|h_{11}|=\sqrt{2}|h_{10}|.
	\end{split}
\end{equation}
Similar expressions to Eq. (\ref{deq:1})-(\ref{deq:11}) can be employed to compute the strong decay widths of the isospin counterparts of baryons outlined in Eq. (\ref{deq:1})-(\ref{deq:11}).
The total strong decay widths computed within this framework are presented in the second column of Table \ref{tab:table0}. By comparing these calculated widths with experimentally measured widths, we assign quantum numbers to the experimentally observed states. The experimental widths  from the PDG \cite{Workman:2022ynf} are listed in the third column alongside the corresponding experimental states in the last column of  Table \ref{tab:table0}. Our results show a good agreement with the experimental widths of $\Xi_{b}$, $\Sigma_{b}$, and $\Xi_{b}^{'}$ baryons. But, for $\Omega_{b}$ baryon, we observe a large strong decay width,   $\Gamma=372.8$ MeV, for the $|1P,1/2^-\rangle_{j=0}$ state. The calculated width in Ref.\cite{Liang:2020hbo,Xiao:2020oif,PhysRevD.101.114013} also shows a large width for this state.
Further, for $\Omega_{b}^{-}|1P,1/2^-\rangle_{j=1}$, $\Omega_{b}^{-}|1P,3/2^-\rangle_{j=1}$ states, $\Xi_{b}^{'}K$ channel is only allowed channel as $\Xi_{b}K$ channel is restricted in heavy quark limit. However, in our model, the $\Xi_{b}^{'}K$ channel is also suppressed due to phase space constraints. In addition, the widths calculated for $\Omega_{b}^{-}|1P,3/2^-\rangle_{j=2}$ and $\Omega_{b}^{-}|1P,5/2^-\rangle_{j=2}$ states are much larger than the widths found for the excited resonances for $\Omega_{b}$ baryons, named $\Omega_{b}(6315)$, $\Omega_{b}(6330)$, $\Omega_{b}(6340)$, and $\Omega_{b}(6350)$ \cite{Workman:2022ynf}. Therefore, it is not possible to attribute any specific spin parity to the four resonances observed for $\Omega_{b}$ baryon. This indicates that additional theoretical and experimental investigation is necessary to identify these states.

\begin{table*}
	\caption{\label{tab:table0}Strong decay widths (in MeV) of singly bottom baryonic states  with available strong decay channels. The strong decay widths are  compared with the experimental widths listed in PDG \cite{Workman:2022ynf} to assign quantum numbers to  observed states.}
	\begin{ruledtabular}
		\begin{tabular}{lccc}

			Decay&Present  &PDG \cite{Workman:2022ynf}&Assignment\\\hline
			
			$\Xi_{b}^{-}$\textit{$|1P,1/2^-\rangle$}$\rightarrow\Xi_{b}^{'-}\pi^{0},\Xi_{b}^{'0}\pi^{-}$         & 3.2                & $<1.9$ & $\Xi_{b}(6100)^{-}$ \\
			$\Xi_{b}^{0}$\textit{$|1P,1/2^-\rangle$}$\rightarrow\Xi_{b}^{'-}\pi^{+},\Xi_{b}^{'0}\pi^{0}$         & 3.2                & &  \\
			
			$\Xi_{b}^{-}$\textit{$|1P,3/2^-\rangle$}$\rightarrow\Xi_{b}^{'-}\pi^{0},\Xi_{b}^{'0}\pi^{-},\Xi_{b}^{*-}\pi^{0},\Xi_{b}^{*0}\pi^{-}$         & 7.6               &   \\
			$\Xi_{b}^{0}$\textit{$|1P,3/2^-\rangle$}$\rightarrow\Xi_{b}^{'0}\pi^{0},\Xi_{b}^{'+}\pi^{-},\Xi_{b}^{*0}\pi^{0},\Xi_{b}^{*+}\pi^{-}$           &  7.6              &   \\
			\hline
			$\Sigma_{b}^{+}$\textit{$|1S,1/2^+\rangle$}$\rightarrow\Lambda_{b}^{0}\pi^{+}$   & 7.1 & 5.0$\pm$0.5 & $\Sigma_{b}^{+}$\\
			$\Sigma_{b}^{0}$\textit{$|1S,1/2^+\rangle$}$\rightarrow\Lambda_{b}^{0}\pi^{0}$   & 7.8 &   \\
			$\Sigma_{b}^{-}$\textit{$|1S,1/2^+\rangle$}$\rightarrow\Lambda_{b}^{0}\pi^{-}$   & 7.1 & 5.3$\pm$0.5 & $\Sigma_{b}^{-}$\\
			$\Sigma_{b}^{+}$\textit{$|1S,3/2^+\rangle$}$\rightarrow\Lambda_{b}^{0}\pi^{+}$  & 12.3 & 9.4$\pm$0.5 & $\Sigma_{b}^{*+}$ \\
			$\Sigma_{b}^{0}$\textit{$|1S,3/2^+\rangle$}$\rightarrow\Lambda_{b}^{0}\pi^{+}$  & 12.3 &   \\
			$\Sigma_{b}^{-}$\textit{$|1S,3/2^+\rangle$}$\rightarrow\Lambda_{b}^{0}\pi^{-}$  & 12.3 & 10.4$\pm$0.8 &  $\Sigma_{b}^{*-}$ \\
			
			$\Sigma_{b}^{+}$\textit{$|1P,1/2^-\rangle_{j=0}$}$\rightarrow\Lambda_{b}^{0}\pi^{+}$   & 288.5 &   \\	
			$\Sigma_{b}^{0}$\textit{$|1P,1/2^-\rangle_{j=0}$}$\rightarrow\Lambda_{b}^{0}\pi^{0}$   & 289.5 &   \\	
			$\Sigma_{b}^{-}$\textit{$|1P,1/2^-\rangle_{j=0}$}$\rightarrow\Lambda_{b}^{0}\pi^{-}$   & 288.5 &   \\	
			$\Sigma_{b}^{+}$\textit{$|1P,1/2^-\rangle_{j=1}$}$\rightarrow\Sigma_{b}^{+}\pi^{0}, \Sigma_{b}^{0}\pi^{+}$   & 93.1  &   \\
			$\Sigma_{b}^{0}$\textit{$|1P,1/2^-\rangle_{j=1}$}$\rightarrow\Sigma_{b}^{0}\pi^{0}, \Sigma_{b}^{+}\pi^{-},\Sigma_{b}^{-}\pi^{+}$   & 139.4 &   \\
			$\Sigma_{b}^{-}$\textit{$|1P,1/2^-\rangle_{j=1}$}$\rightarrow\Sigma_{b}^{-}\pi^{0}, \Sigma_{b}^{0}\pi^{-} $  &  93.1  &   \\		
			$\Sigma_{b}^{+}$\textit{$|1P,3/2^-\rangle_{j=1}$}$\rightarrow\Sigma_{b}^{*+}\pi^{0}, \Sigma_{b}^{*0}\pi^{+}$   & $<$ 24.6 & 31$\pm$6 & $\Sigma_{b}(6097)^{+}$   \\
			$\Sigma_{b}^{0}$\textit{$|1P,3/2^-\rangle_{j=1}$}$\rightarrow\Sigma_{b}^{*0}\pi^{0}, \Sigma_{b}^{*+}\pi^{-},\Sigma_{b}^{*-}\pi^{+}$   & $<$ 36.6 &   \\
			$\Sigma_{b}^{-}$\textit{$|1P,3/2^-\rangle_{j=1}$}$\rightarrow\Sigma_{b}^{*-}\pi^{0}, \Sigma_{b}^{*0}\pi^{-} $  &  $<$ 24.6  &  29$\pm$4 & $\Sigma_{b}(6097)^{-}$   \\	
			
			\hline
			
			$\Xi_{b}^{'-}$\textit{$|1S,3/2^+\rangle$}$\rightarrow\Xi_{b}^{-}\pi^{0},\Xi_{b}^{0}\pi^{-}$         & 1.0                & 1.6$\pm$0.33 & $\Xi_{b}(5955)^{-}$ \\	 
			$\Xi_{b}^{'0}$\textit{$|1S,3/2^+\rangle$}$\rightarrow\Xi_{b}^{0}\pi^{0},\Xi_{b}^{-}\pi^{+}$         & 1.0                & 0.9$\pm$0.18 & $\Xi_{b}(5945)^{0}$ \\
			$\Xi_{b}^{'-}$\textit{$|1P,1/2^-\rangle_{j=0}$}$\rightarrow\Xi_{b}^{-}\pi^{0},\Xi_{b}^{0}\pi^{-}$         &     174.9          &   \\
			$\Xi_{b}^{'0}$\textit{$|1P,1/2^-\rangle_{j=0}$}$\rightarrow\Xi_{b}^{0}\pi^{0},\Xi_{b}^{-}\pi^{+}$         &     174.9          &   \\
			
			$\Xi_{b}^{'-}$\textit{$|1P,1/2^-\rangle_{j=1}$}$\rightarrow\Xi_{b}^{'-}\pi^{0},\Xi_{b}^{'0}\pi^{-}$         &      42.3           &   \\
			$\Xi_{b}^{'0}$\textit{$|1P,1/2^-\rangle_{j=1}$}$\rightarrow\Xi_{b}^{'0}\pi^{0},\Xi_{b}^{'-}\pi^{+}$         &      42.3           &   \\
			$\Xi_{b}^{'-}$\textit{$|1P,3/2^-\rangle_{j=1}$}$\rightarrow\Xi_{b}^{'-}\pi^{0},\Xi_{b}^{'0}\pi^{-}$         &     $<$ 21.5           & 19.9$\pm$2.6 & $\Xi_{b}(6227)^{-}$  \\			
			$\Xi_{b}^{'0}$\textit{$|1P,3/2^-\rangle_{j=1}$}$\rightarrow\Xi_{b}^{'0}\pi^{0},\Xi_{b}^{'-}\pi^{+}$         &     $<$ 21.5          & 19$^{+5}_{-4}$ & $\Xi_{b}(6227)^{0}$ \\
			
			\hline
			
			$\Omega_{b}^{-}|1P,1/2^-\rangle_{j=0}\rightarrow\Xi_{b}^{-}K^{0},\Xi_{b}^{0}K^{-}$ &372.8\\
			$\Omega_{b}^{-}|1P,1/2^-\rangle_{j=1}\rightarrow\Xi_{b}^{'-}K^{0},\Xi_{b}^{'0}K^{-}$&0\\
			$\Omega_{b}^{-}|1P,3/2^-\rangle_{j=1}\rightarrow\Xi_{b}^{'-}K^{0},\Xi_{b}^{'0}K^{-}$
			&0\\
			$\Omega_{b}^{-}|1P,3/2^-\rangle_{j=2}\rightarrow\Xi_{b}^{-}K^{0},\Xi_{b}^{0}K^{-},\Xi_{b}^{'-}K^{0},\Xi_{b}^{'0}K^{-}$
			&$<$965.65\\
			$\Omega_{b}^{-}|1P,5/2^-\rangle_{j=2}\rightarrow\Xi_{b}^{-}K^{0},\Xi_{b}^{0}K^{-},\Xi_{b}^{'-}K^{0},\Xi_{b}^{'0}K^{-}$
			&$<$1228.83\\			
		\end{tabular}  
	\end{ruledtabular}
\end{table*}

\section{RESULTS AND DISCUSSION}
Tables \ref{tab:table2}-\ref{tab:table6} display the mass spectra for $\Lambda_{b}$, $\Xi_{b}$, $\Sigma_{b}$, $\Xi_{b}^{'}$, and $\Omega_{b}$ baryons, respectively. In the quark-diquark picture of singly bottom baryons, the possible states and their quantum numbers $ (n, L, J, j)$ are given in the second and first columns, respectively. Then we list the calculated masses for these states in the third column. These calculated masses are then compared with the masses of experimentally observed states, as listed in PDG\cite{Workman:2022ynf} in the fourth column. We also show the mass predictions from other theoretical models in the subsequent columns for comparison. 

\begin{figure*}
	\begin{subfigure}{.5\textwidth}
		\centering
		\includegraphics[width=.9\linewidth]{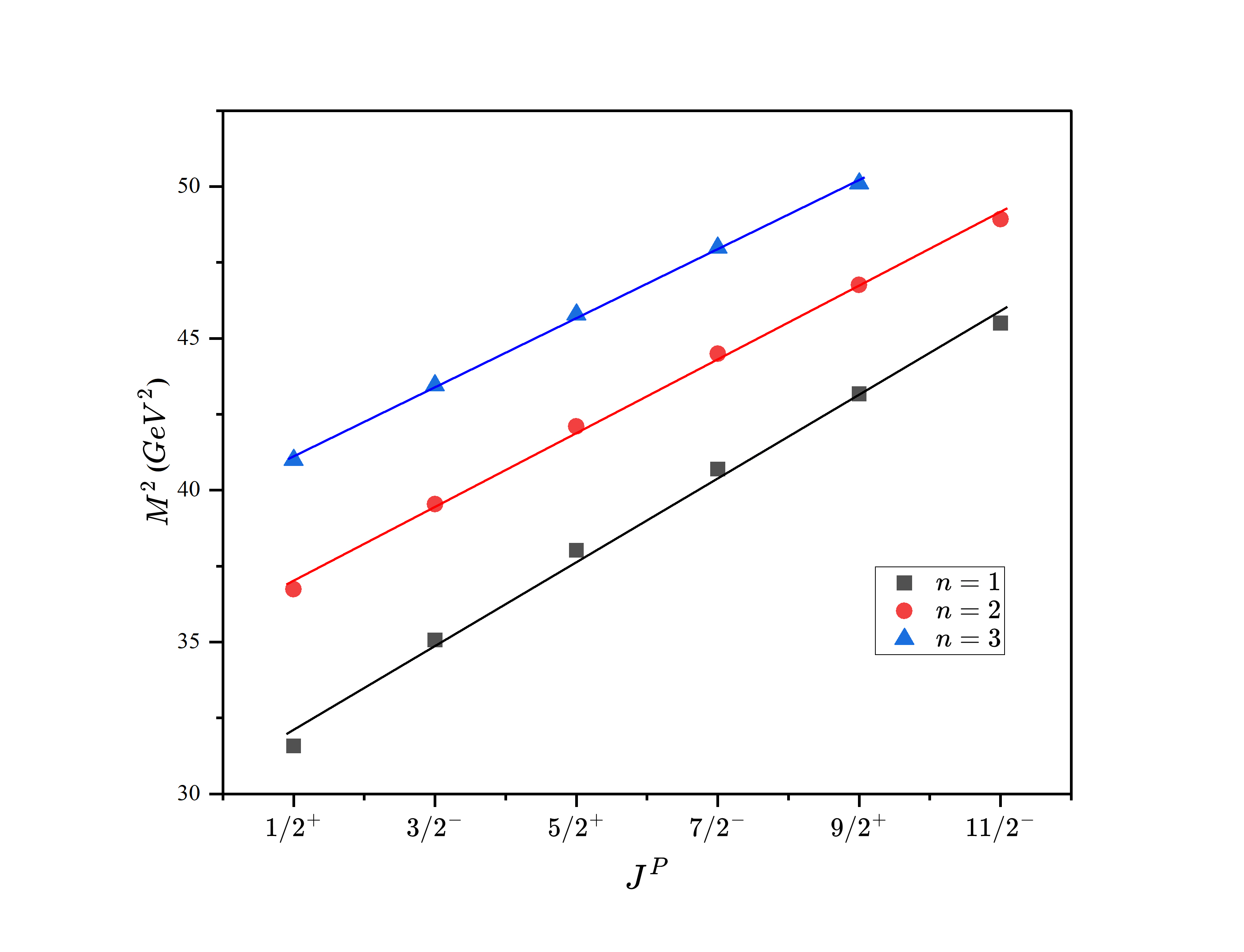}
	\end{subfigure}%
	\begin{subfigure}{.5\textwidth}
		\centering
		\includegraphics[width=.9\linewidth]{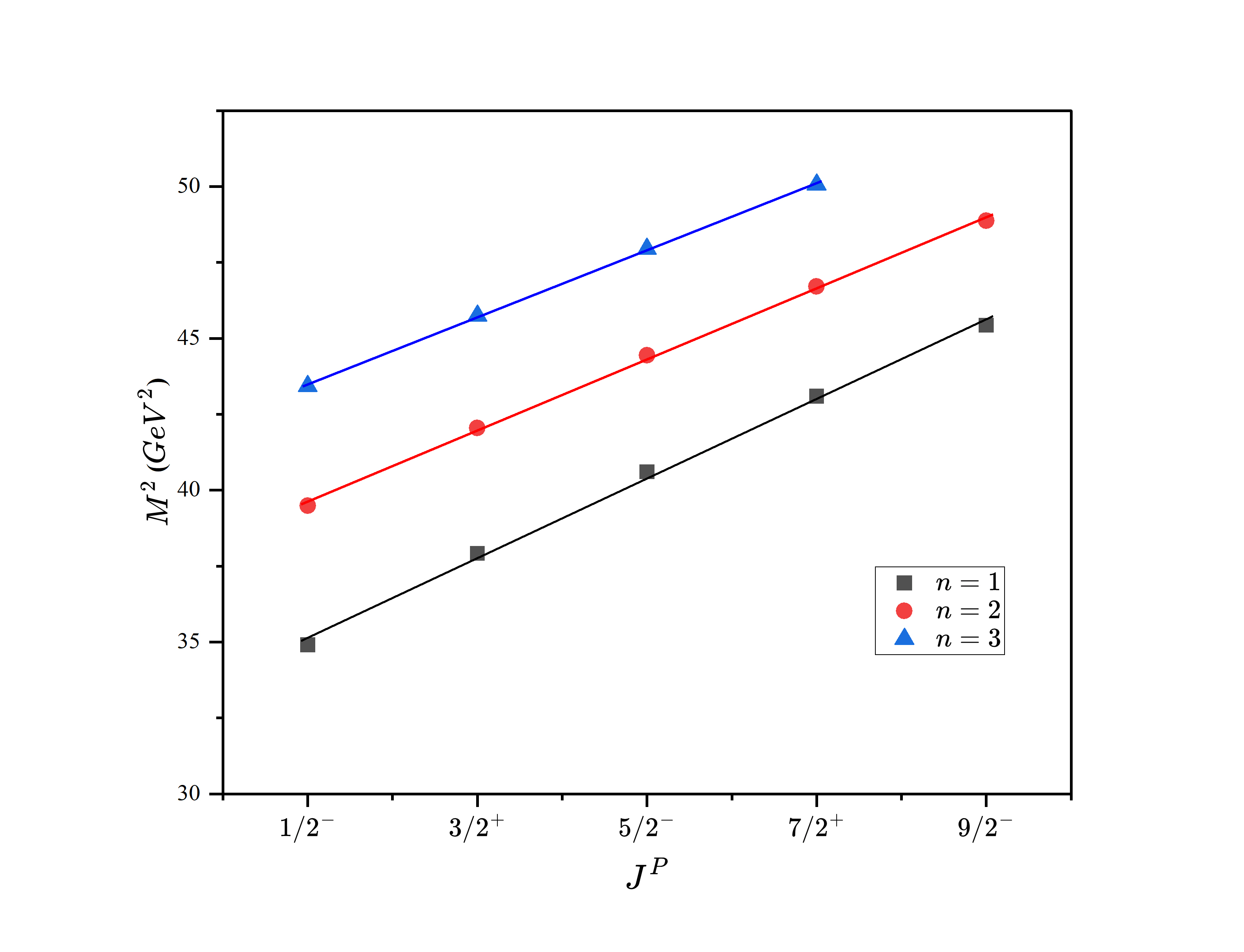}
	\end{subfigure}
	\caption{\label{fig:3}Regge trajectory in the ($J,M^{2}$) plane for $\Lambda_{b}$ baryonic family with natural parity states (left) and unnatural parity states (right).}
\end{figure*}

\begin{figure*}
	\begin{subfigure}{.5\textwidth}
		\centering
		\includegraphics[width=.9\linewidth]{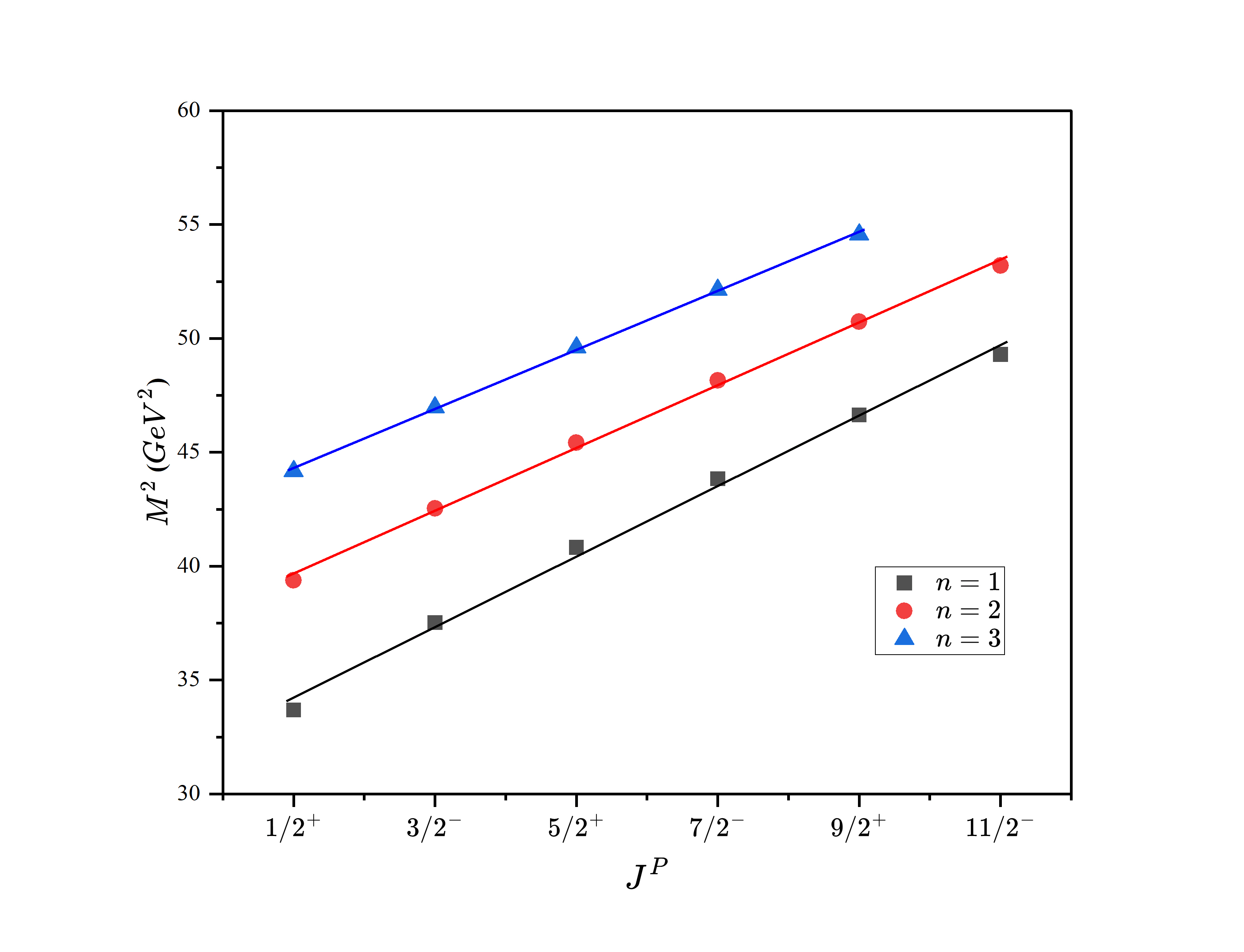}
	\end{subfigure}%
	\begin{subfigure}{.5\textwidth}
		\centering
		\includegraphics[width=.9\linewidth]{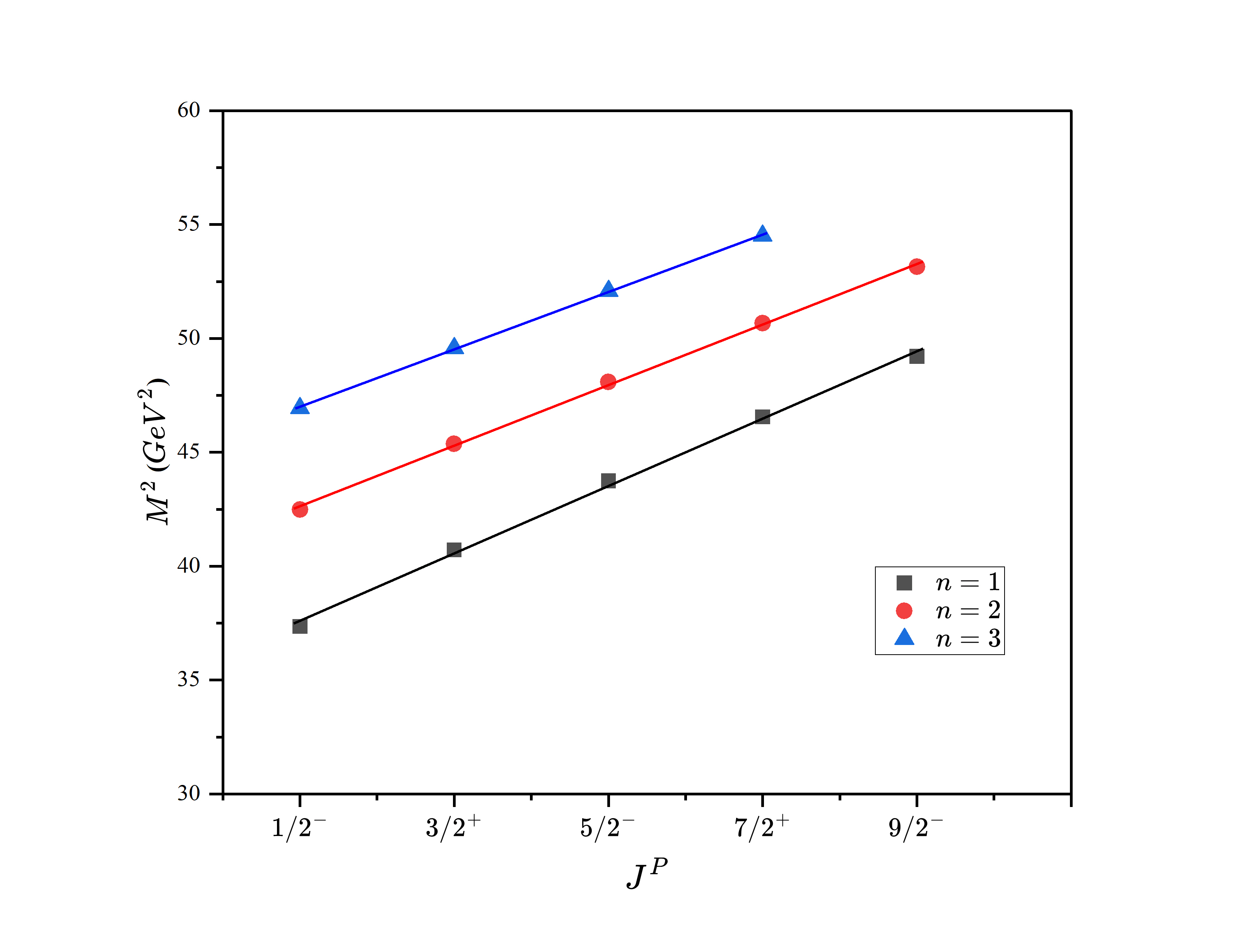}
	\end{subfigure}
	\caption{\label{fig:4}Regge trajectory in the ($J,M^{2}$) plane for $\Xi_{b}$ baryonic family with natural parity states (left) and unnatural parity states (right).}
\end{figure*}

\begin{figure*}
	\begin{subfigure}{.5\textwidth}
		\centering
		\includegraphics[width=.9\linewidth]{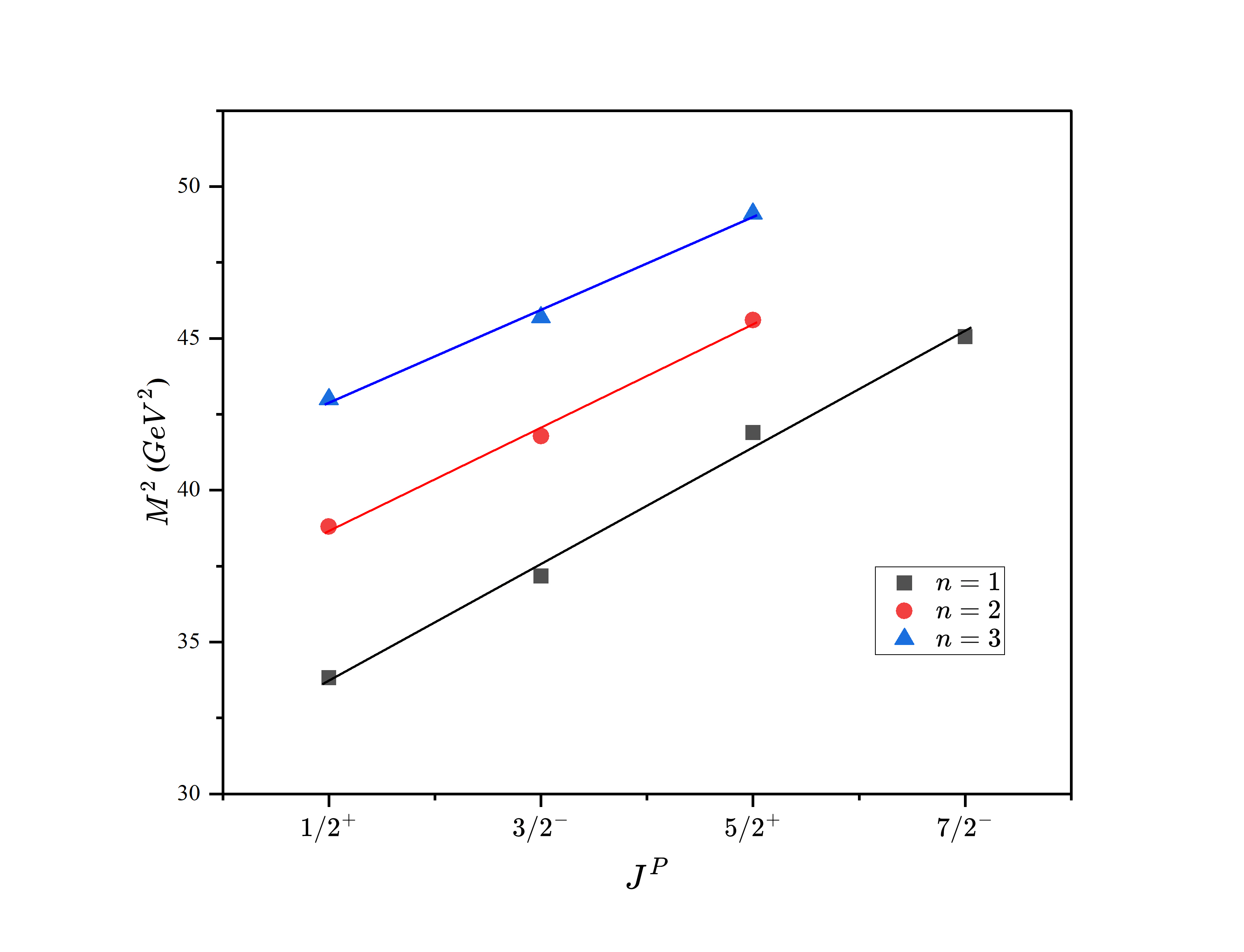}
	\end{subfigure}%
	\begin{subfigure}{.5\textwidth}
		\centering
		\includegraphics[width=.9\linewidth]{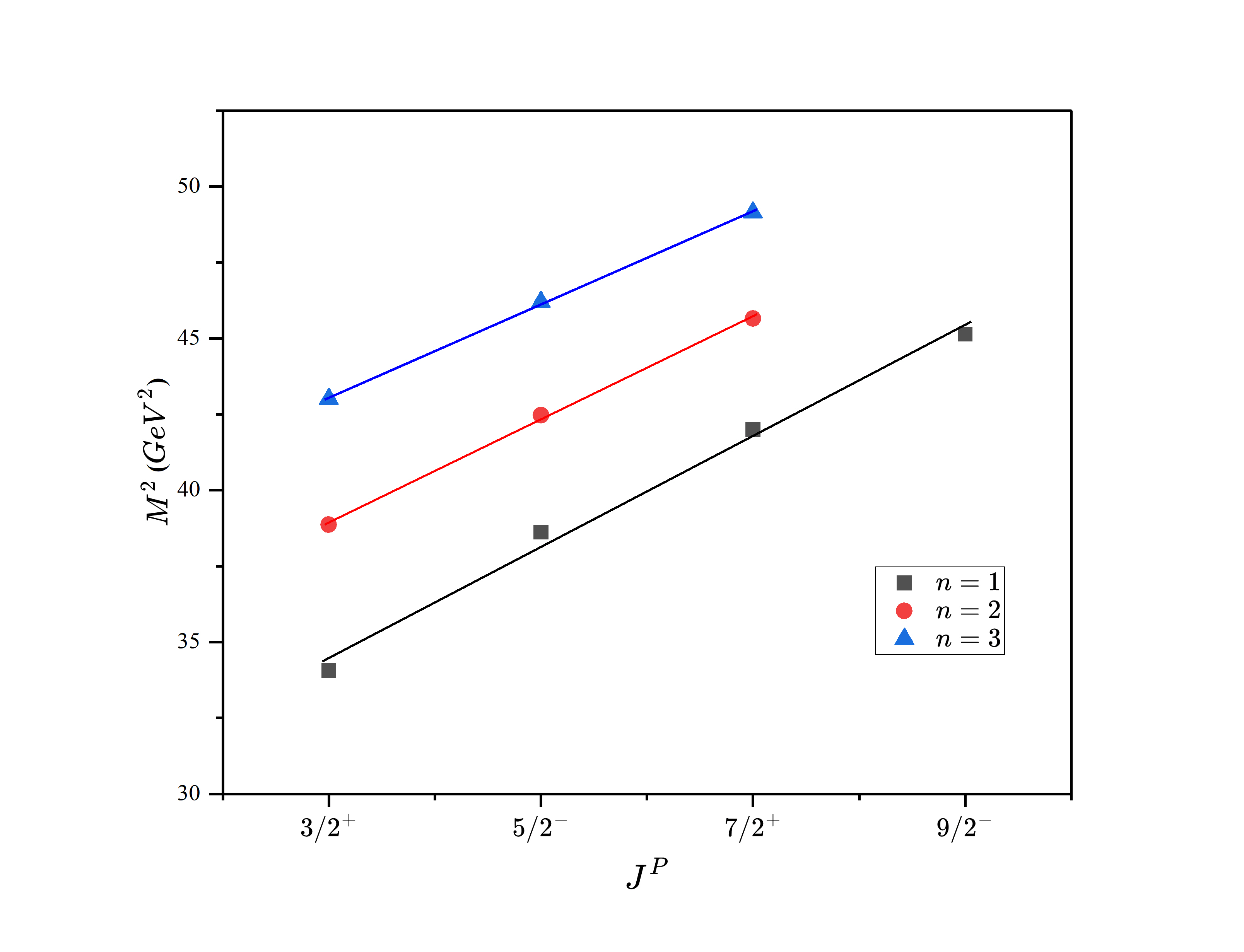}
	\end{subfigure}
	\caption{\label{fig:5}Regge trajectory in the ($J,M^{2}$) plane for $\Sigma_{b}$ baryonic family with natural parity states (left) and unnatural parity states (right).}
\end{figure*}

\begin{figure*}
	\begin{subfigure}{.5\textwidth}
		\centering
		\includegraphics[width=.9\linewidth]{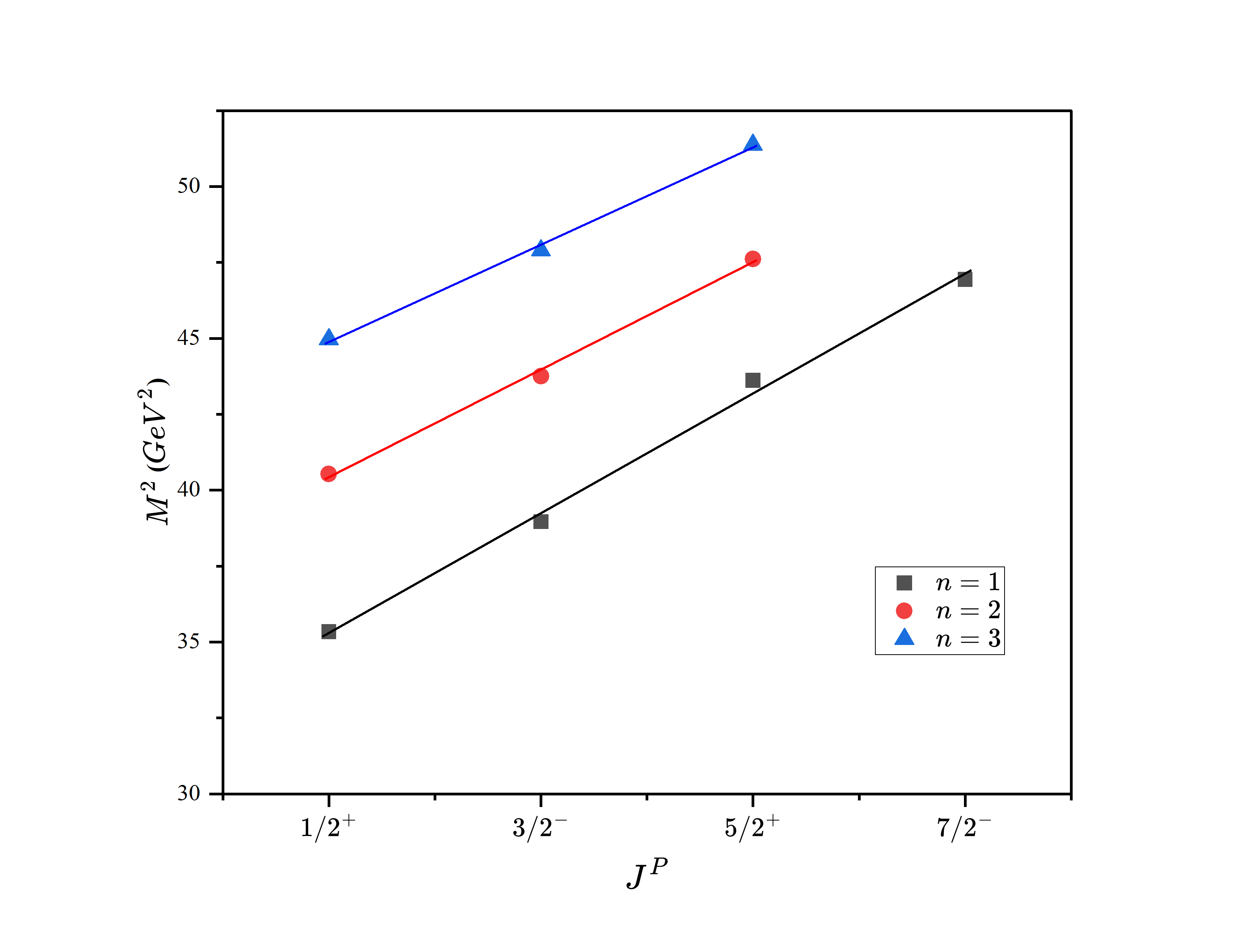}
	\end{subfigure}%
	\begin{subfigure}{.5\textwidth}
		\centering
		\includegraphics[width=.9\linewidth]{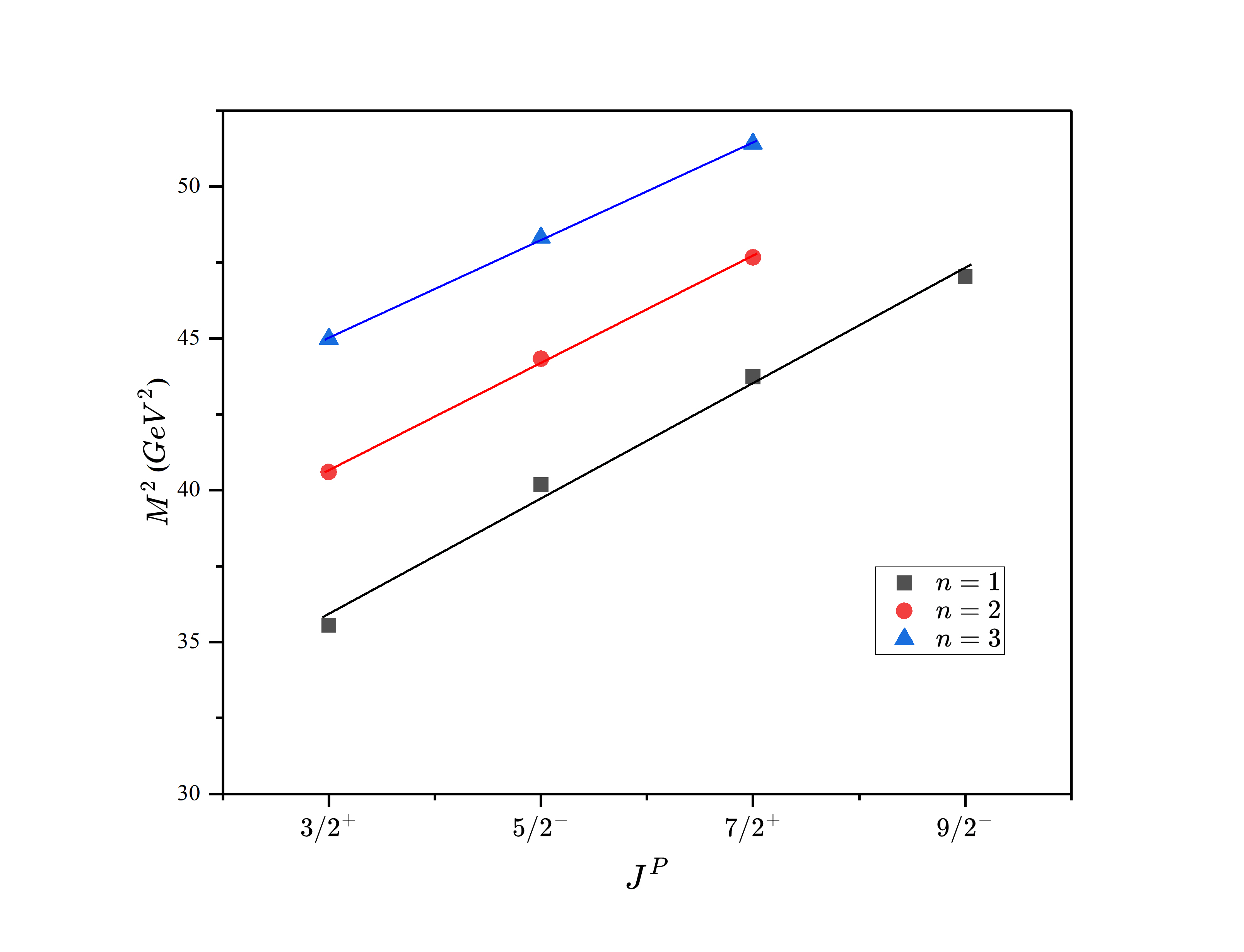}
	\end{subfigure}
	\caption{\label{fig:6}Regge trajectory in the ($J,M^{2}$) plane for $\Xi_{b}^{'}$ baryonic family with natural parity states (left) and unnatural parity states (right).}
\end{figure*}

\begin{figure*}
	\begin{subfigure}{.5\textwidth}
		\centering
		\includegraphics[width=.9\linewidth]{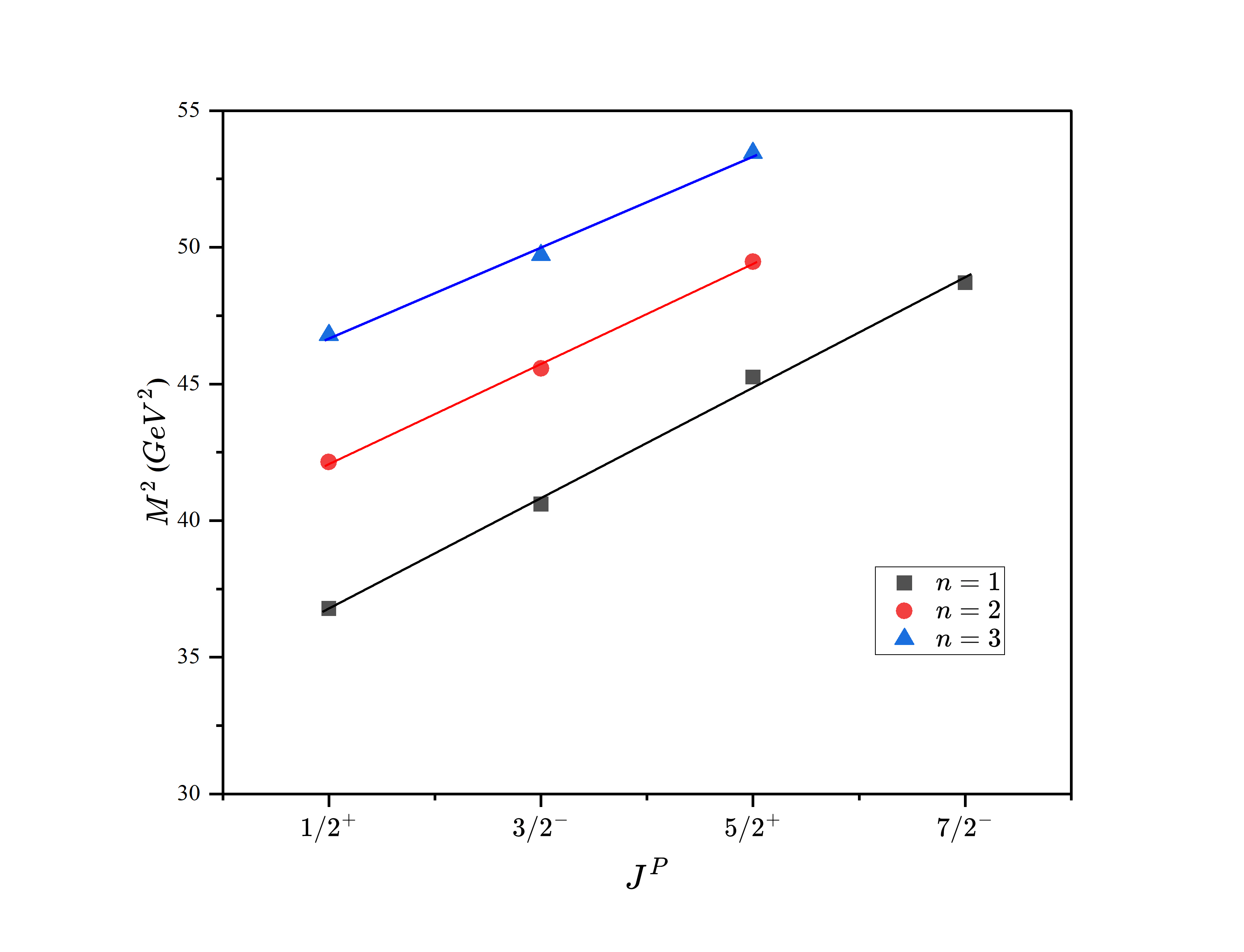}
	\end{subfigure}%
	\begin{subfigure}{.5\textwidth}
		\centering
		\includegraphics[width=.9\linewidth]{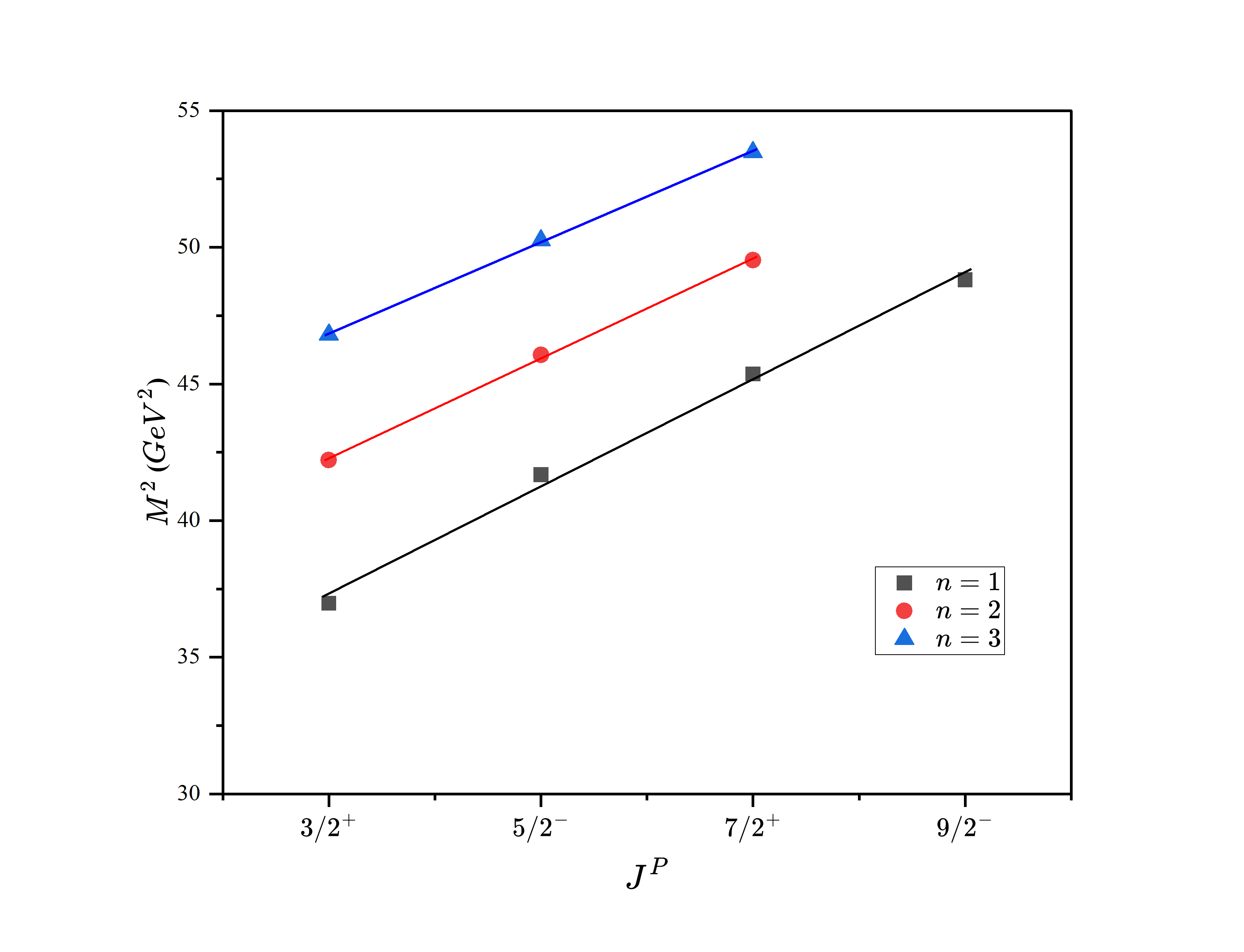}
	\end{subfigure}
	\caption{\label{fig:7}Regge trajectory in the ($J,M^{2}$) plane for $\Omega_{b}$ baryonic family with natural parity states (left) and unnatural parity states (right).}
\end{figure*}

Further, using the calculated mass spectra of singly bottom baryons, we analyse the Regge trajectories in the $(J,M^{2})$ plane for natural and unnatural parity states, as shown in Figs. \ref{fig:3}-\ref{fig:7}. Each line in these figures corresponds to a different principle quantum number: n = 1, 2, and 3. The computed masses fit well with linear trajectories. Moreover, the Regge trajectories are almost parallel and evenly spaced.    

In the next part of this section, we compare our theoretical results with the experimental data to assign possible spin-parity  quantum numbers to bottom baryons reported in PDG. Assigning spin parity to states observed experimentally becomes more reliable when conducted through a combination of mass spectra examination and decay analysis. Initially, we employ our mass spectrum to identify resonances of bottom baryons, considering it as the primary factor, with decay width serving as a secondary factor. In instances where the mass spectra suggest multiple assignments, we will turn to decay width calculations to eliminate certain spin-parity possibilities inferred from the mass spectra.

\subsection{$\Lambda_{b}$ baryons}
For the $\Lambda_{b}$ baryonic family, the states belonging to $1S$, $1P$, and $1D$ wave have been well established. The two narrow $1P$-wave $\Lambda_{b}$ baryons, denoted as $\Lambda_{b}(5912)$ and  $\Lambda_{b}(5920)$, were first discovered by the LHCb Collaboration in 2012 in the $\Lambda_{b}^{0}\pi^{+}\pi^{-}$ spectrum \cite{PhysRevLett.109.172003}. They were later confirmed by the CDF collaboration, \cite{PhysRevD.88.072003}. Masses of these states are well reproduced in our model. They also match well with Refs. \cite{Ebert:2011kk, YU2023116183, Oudichhya:2021yln, article, Garcia-Tecocoatzi:2023btk}. Since the strong decay channel is not available for $1S$- and $1P$-wave states of the $\Lambda_{b}$ baryon, their strong decay widths are equal to zero.
Recently, in 2020, the two $1D$ wave $\Lambda_{b}$ candidates, $\Lambda_{b}^{0}(6146)$ and  $\Lambda_{b}^{0}(6152)$, were also discovered by LHCb in the $\Lambda_{b}^{0}\pi^{+}\pi^{-}$ spectrum  \cite{PhysRevLett.123.152001}. Here we observe that the experimental masses of these two states are also very close to our prediction for the two states of the $1D$-wave, with differences of 11.5 MeV and 13.7 MeV, respectively. Here, note that the experimental masses of the $1S$- and $1P$-waves of the $\Lambda_{b}$ baryonic system are used as inputs to determine the parameters $m_{b}$ and $\sigma_{\Lambda_{b}}$. We have now calculated the masses of the states belonging to the $1D$-wave using these parameters, which agrees well with the experimentally known states of the $1D$-wave. Furthermore, estimated masses for the $1D$-wave display a good agreement with the results reported in Refs. \cite{article,YU2023116183}, whereas the predictions made in studies \cite{Ebert:2011kk, Garcia-Tecocoatzi:2023btk, Oudichhya:2021yln} are found to be overestimated.

It is also worth mentioning the recent discovery of the $\Lambda_{b}(6070)$ state by LHCb and CMS experiments \cite{LHCb:2020lzx, 2020135345} in the $\Lambda_{b}^{0}\pi^{+}\pi^{-}$ channel. This state is established to be the first radial excitation ($2S$) of $\Lambda_{b}$ baryon \cite{Workman:2022ynf}. We observe that it's experimental mass is in good agreement with the calculated mass for the first radial excitation ($2S$) with a slight difference of 11.3 MeV only. While theoretical studies in Refs. \cite{Ebert:2011kk, YU2023116183, Oudichhya:2021yln, article, Garcia-Tecocoatzi:2023btk} show a deviation of 28–172 MeV. To calculate this radially excited state, we have used the value of parameter $\lambda$, which is extracted from the experimental data of singly charmed baryons in our previous work \cite{PhysRevD.108.014011}. A close match between the experimental mass and the calculated mass of $|2S, 1/2^{+}\rangle$ state suggests that the parameters extracted from the singly charmed baryons are also able to explain masses of singly bottom baryons, and our predictions for further excited states are reliable.

\subsection{$\Xi_{b}$ and $\Xi_{b}^{'}$ baryons}
For the $\Xi_{b}$ baryonic family, only the state belonging to ground state $|1S, 1/2^{+}\rangle$ is established. Our calculated mass for this ground state is in good agreement with it, with a difference of only 6.6 MeV. The strong decay is forbidden for the ground state of $\Xi_{b}$ baryon.
Recently, in 2021, the CMS experiment reported the $\Xi_{b}(6100)$ state in the $\Xi_{b}^{-}\pi^{+}\pi^{-}$ channel \cite{CMS:2021rvl}. Its spin parity quantum numbers are not measured in this experiment. But based on similarities with known excited $\Xi_{c}$ baryon states, they predicted this state to be the orbitally excited $\Xi_{b}$ baryon, with spin-parity $J^{P}=\frac{3}{2}^{-}$. However, this spin parity has not yet been verified.
Our calculated mass for the $|1P, 1/2^{-}\rangle$ state falls very close to the experimentally measured mass of the $\Xi_{b}(6100)$ state, with a difference of only 11.5 MeV.
Furthermore, as indicated in Table \ref{tab:table0}, the decay width of the $\Xi_{b}(6100)$ state closely aligns with the calculated strong decay width of the $|1P, 1/2^{-}\rangle$ state compared to the calculated decay width of the $|1P, 3/2^{-}\rangle$ state.
Hence, we predict $\Xi_{b}(6100)$ state to be a good candidate of $1P$-wave with spin-parity $J^{P}=\frac{1}{2}^{-}$. The mass prediction in Ref.\cite{Ebert:2011kk} is in favour of this assignment.

For the $\Xi_{b}^{'}$ baryonic family, the states belonging to the $1S$-wave, with $J^{P}=\frac{1}{2}^{+}$ and $J^{P}=\frac{3}{2}^{+}$, are well determined. Our calculated masses for these states are very close to the experimental masses, with a difference of only 10 MeV and 7.3 MeV, respectively.
The strong decay of the ground state of $\Xi_{b}^{'}$ baryon is forbidden. But, our calculated strong decay width for $\Xi_{b}^{'-}|1S, 3/2^{+}\rangle$ and $\Xi_{b}^{'0}|1S, 3/2^{+}\rangle$ states is compatible with the decay width of the observed states $\Xi_{b}(5955)^{-}$ and $\Xi_{b}(5945)^{0}$, respectively.
Further, the $\Xi_{b}(6227)^{-}$ state was observed by the LHCb experiment in the $\Xi_{b}^{0}\pi^{-}$ channel \cite{PhysRevLett.121.072002}. Later, the LHCb experiment in the $\Xi_{b}^{-}\pi^{+}$ channel observed its isospin partner named $\Xi_{b}(6227)^{0}$. The spin-parity quantum numbers of this state are not yet identified.
Our calculated masses for $|1P, 1/2^{-}\rangle_{j=1}$ and $|1P, 3/2^{-}\rangle_{j=1}$ states of $\Xi_{b}^{'}$ baryon   are  close to the experimental mass of the $\Xi_{b}(6227)$ baryon. However, the decay width of the $\Xi_{b}(6227)$ state is nearly identical to the strong decay width of the $|1P, 3/2^{-}\rangle_{j=1}$ state, which eliminates the possibility of it being the $|1P, 1/2^{-}\rangle_{j=1}$ state.
Therefore, we identify the $\Xi_{b}(6227)$ baryon as the first orbital excitation ($1P$) of $\Xi_{b}^{'}$ baryon with  $J^{P} = \frac{3}{2}^{-}$.
This assignment is confirmed by Ref.\cite{Oudichhya:2021yln}. While in Ref. \cite{Ebert:2011kk, li2022systematic} the predicted masses of states in the $1P$-wave of $\Xi_{b}^{'}$ baryon are too close to assign specific spin-parity to $\Xi_{b}(6227)$. However, these predictions do provide support for our argument that the $\Xi_{b}(6227)$ belongs to the $1P$-wave of the $\Xi_{b}^{'}$ baryon. The authors in Ref.\cite{Garcia-Tecocoatzi:2023btk} also support this argument, but they predict it's spin-parity to be $J^{P}=\frac{5}{2}^{-}$.
At last, in 2021, the LHCb collaboration has observed $\Xi_{b}(6327)$ and $\Xi_{b}(6333)$ states in the $\Lambda_{b}^{0}K^{-}\pi^{+}$ channel \cite{PhysRevLett.128.162001}. The spin-parity of it is still a mystery. The masses and decay widths of these states were in agreement with the predictions made in Ref.\cite{PhysRevD.100.094032,PhysRevD.98.076015} for $1D$-wave $\Xi_{b}$ baryonic states with $J^{P}= \frac{3}{2}^{-}$ and $\frac{5}{2}^{-}$. However, we observe that the masses of $\Xi_{b}(6327)$ and $\Xi_{b}(6333)$ are very close to our predicted masses for the $|1P, 3/2^{-}\rangle_{j=2}$ and $|1P, 5/2^{-}\rangle$ states of the $\Xi_{b}^{'}$ baryonic family. Hence, we suggest the alternative possibility that $\Xi_{b}(6327)$ and $\Xi_{b}(6333)$ can be components of the first orbital excitation ($1P$) of the $\Xi_{b}^{'}$ baryonic family, with $J^{P}= \frac{3}{2}^{-}$ and $\frac{5}{2}^{-}$, respectively. 

\subsection{$\Sigma_{b}$ baryons}
The $\Sigma_{b}$ baryonic family has very well established states in the $1S$ wave. Our calculated mass for the $1S$ wave with quantum numbers $J^{P}=\frac{1}{2}^{+}$ and $J^{P}=\frac{3}{2}^{+}$ is very close to the experimental masses of these states, with a difference of 5.64 MeV and 6.68 MeV only.
As indicated in Table \ref{tab:table0}, their widths are also well reproduced in our model.

Further, the LHCb Collaboration has been able to detect only one excited state of the $\Sigma_{b}$ baryon so far, named $\Sigma_{b}(6097)$ in the $\Lambda_{c}^{+}K^{-}\pi^{+}\pi^{-}$ mass spectrum \cite{PhysRevLett.122.012001}. The quantum numbers of this state are not confirmed yet. The experimental mass of this state is in excellent match with our predicted mass for the $|1P, 3/2^-\rangle_{j=1}$ state.
In addition, the decay width of $\Sigma_{b}(6097)$ also closely matches the strong decay width of $|1P, 3/2^-\rangle_{j=1}$ state.
This clearly indicates that $\Sigma_{b}(6097)$ is a potential candidate for a $1P$- wave with a spin-parity  $J^{P}=\frac{3}{2}^{-}$. This assignment is also supported by ref. \cite{Oudichhya:2021yln, PhysRevD.99.094003, PhysRevD.98.074032}. 

\subsection{$\Omega_{b}$ baryons}
For $\Omega_{b}$ baryons, only the ground state is established. Our calculated mass for the ground state deviates slightly by 20 MeV from its experimental mass. \textbf{Further, the LHCb collaboration \cite{PhysRevLett.124.082002} has reported four narrow states in the $\Xi_{b}^{0}K^{-}$ spectrum, designated as $\Omega_{b}(6316)$, $\Omega_{b}(6330)$, $\Omega_{b}(6340)$, and $\Omega_{b}(6350)$. These states have masses ranging from 6315 to 6350 MeV, though their spin parity remains unmeasured. Similarly, the LHCb  collaboration \cite{LHCb:2017uwr} has observed five narrow states of the $\Omega_{c}$ baryonic family, named $\Omega_{c}(3000)$,  $\Omega_{c}(3050)$,  $\Omega_{c}(3065)$,  $\Omega_{c}(3090)$, and $\Omega_{c}(3120)$. In our previous work \cite{PhysRevD.108.014011}, we have interpreted them as the $1P$-wave excitations of $\{s,s\}$ diquark with respect to the charm quark. The discovery of these narrow states in both the $\Omega_{b}$ and $\Omega_{c}$ families reinforces the notion of similar excitation mechanisms in doubly strange baryons containing one heavy quark.}

As shown in Table \ref{tab:table6}, the masses of four experimentally observed excited states of $\Omega_{b}$ baryon lie comparatively close to the masses of the $1P$-wave states, named $|1P, 1/2^-\rangle_{j=0}$, $|1P, 1/2^-\rangle_{j=1}$, $|1P, 3/2^-\rangle_{j=1}$, and $|1P, 3/2^-\rangle_{j=2}$. Therefore, it is possible that these four narrow states belong to the $1P$-wave. However, in our calculated mass spectra, the mass splitting in the $1P$-wave is large, resulting in a considerable difference between the calculated masses of the $1P$-wave states and the masses of experimentally detected states. A potential reason for large splitting in $1P$-wave could be our underlying assumption that the $\Omega_{b}$ baryon follows the heavy quark symmetry. This assumption has led us to calculate the spin-dependent splitting in the j-j coupling scheme. However, it may not be the most effective coupling scheme for the $\Omega_{b}$ baryon, which consists of two strange quarks. This suggests that an alternative coupling scheme must be developed in order to study the mass spectra of $\Omega_{b}$ baryons in the relativistic flux tube model.

Based on calculation of strong decay widths, we cannot definitively determine the spin-parity assignments for the observed excited states of the $\Omega_{b}$ baryon. The calculated widths for $|1P, 1/2^-\rangle_{j=0}$ is too broad to be observed, while the range of widths for $|1P, 3/2^-\rangle_{j=1}$ and $|1P, 3/2^-\rangle_{j=2}$ are too large to firmly associate any experimental state with them. Additionally, for the remaining two states in the $1P$-wave, strong decay channels are suppressed by phase space constraint in our model. Hence, further theoretical and experimental research is required to  identify these four resonances of $\Omega_{b}$ baryon.

\section{CONCLUSION AND OUTLOOK}
We have conducted computations for the mass spectra of single-bottom baryons. In the relativistic flux tube model, single-bottom baryons are pictured as a two-body system consisting of a bottom quark and a diquark. The additional spin-dependent interactions are also taken into account in the j-j coupling scheme. It is essential to mention that our approach does not include parameter fitting. Instead, the parameters of our model were determined earlier by using the masses of experimentally reported states of singly charmed baryons and some low lying states of $\Lambda_b$ baryon as inputs. Therefore, it provide a unified description of both singly charmed and bottom baryons. In various quark model calculations \cite{Ebert:2011kk,Garcia-Tecocoatzi:2023btk}, model parameters are typically adjusted to fit the ground state or a lower excited state, resulting in a close match to experimentally measured masses for the ground state. However, In that approach the discrepancies between model predictions and experimental data become more pronounced for higher excited states such as $1D$-wave states of $\Lambda_{b}$ baryon and $1P$-wave states of $\Xi_{b}$ baryon. While in our  approach, we found that the calculated masses for excited states of $\Lambda_{b}$, $\Xi_{b}$, $\Sigma_{b}$,and $\Xi_{b}^{'}$ baryons are closer to the experimentally measured masses. Therefore, our model prediction for the excited state masses are more reliable.
The Regge trajectories obtained from the calculated masses are observed to be almost linear, parallel, and equidistant in the $(J,M^{2})$ plane. 
The masses and strong decay width of experimentally observed states of $\Lambda_{b}$, $\Xi_{b}$, $\Sigma_{b}$, and $\Xi_{b}'$ are well reproduced in this theoretical framework. Based on this, we have assigned possible spin-parity quantum numbers to $\Sigma_{b}(6097)$, $\Xi_{b}(6100)$, $\Xi_{b}(6227)$, $\Xi_{b}(6327)$, and $\Xi_{b}(6333)$, which might be useful to establish them in their mass spectra. 
Ongoing experimental research on single-bottom baryons at LHCb is expected to discover new states with increasing luminosity and energy in the future. Our theoretical predictions for the masses of higher-lying states can assist experimentalists in identifying and investigating certain resonances in the spectrum of singly bottom baryons.

For the $\Omega_{b}$ baryonic family, as the experimentally measured masses of $\Omega_{b}(6316)$, $\Omega_{b}(6330)$, $\Omega_{b}(6340)$, and $\Omega_{b}(6350)$ fall close to the theoretical masses of $1P$-wave states in the quark-diquark picture. But, with the j-j coupling scheme in our model, we observe significant spin-dependent mass splitting in the $1P$-wave. To accurately reproduce the experimental masses of $\Omega_b$ baryons, a different coupling scheme needs to be introduced. In the future, it would be interesting to explore an alternative coupling scheme for studying the mass spectra of $\Omega_{b}$ baryons in the relativistic flux tube model.

\begin{acknowledgments}
	Ms. Pooja Jakhad acknowledges the financial assistance by the Council of Scientific \& Industrial Research (CSIR) under the JRF-FELLOWSHIP scheme with file no. 09/1007(13321)/2022-EMR-I. 
	
\end{acknowledgments}

 \appendix
 
  \section*{Appendix}
 
 \begin{widetext}
 	
 	We have already demonstrated how to compute the expectation values of operators involved in  spin dependent interactions for  $S$-wave, $P$-wave, and $D$-wave  of singly heavy baryons having vector diquarks in our earlier work \cite{PhysRevD.108.014011}. In this section, we will extend our calculation to the case of $F$-wave and $G$-wave of singly heavy baryons having vector diquark.
 	
 	\begin{enumerate}
 		\item{\textit{The F-wave}}:
 		
 		The  $L-S$  coupling scheme involves coupling of $\mathbf{S_{\mathcal{D}}}$ and $\mathbf{S_{b}}$ to form $\mathbf{S}$, and then coupling $\mathbf{S}$ and $\mathbf{L}$ to get the total angular momentum $\mathbf{J}$. Using uncoupled states $ |S_{\mathcal{D}}, S_{\mathcal{D}_{3}}\rangle $, $ |S_{b}, S_{b_{3}}\rangle $, and  $ |L, L_{3}\rangle $ , we can construct the  $L-S$  coupling basis states as  \cite{PhysRevD.108.014011}
 		
 		\begin{equation}\label{B1}
 			|(S_{\mathcal{D}}S_{b   })SL;J\ J_{3}\rangle= \displaystyle\sum_{\scriptscriptstyle {S_{\mathcal{D}_{3}}S_{b   _{3}}L_{3}}S_{3}} C_{S_{\mathcal{D}_{3}}S_{b   _{3}}S_{3}}^{S_{\mathcal{D}}\ S_{b   }\ S}\ 
 			C_{S_{3}L_{3}J_{3}}^{S\ L\ J}\ |S_{\mathcal{D}}S_{\mathcal{D}_{3}}\rangle |S_{b}S_{b_{3}}\rangle  |L\ L_{3}\rangle,
 		\end{equation}
 		where, $ S_{\mathcal{D}_{3}} $, $ S_{b_{3}} $, $ L_{3} $ and $ J_{3} $ signify the third component of $\mathbf{S_{\mathcal{D}}}$, $\mathbf{S_{b}}$, $\mathbf{L}$ and $\mathbf{J}$, respectively. ${ C_{S_{\mathcal{D}_{3}}S_{b_{3}}S_{3}}^{S_{\mathcal{D}}\ S_{b}\ S}}$ and $ C_{S_{3}L_{3}J_{3}}^{S\ L\ J}$ are Clebsch-Gordan coefficients.
 		For the sake of convenience, We use $ |^{2S+1}L_{J};J_{3}\rangle $ to denote the basis $ |(S_{\mathcal{D}}S_{b})SL;J\ J_{3}\rangle $ and $|S_{\mathcal{D}_{3}},S_{b_{3}},L_{3}\rangle$ to denote the product of states $|S_{\mathcal{D}}S_{\mathcal{D}_{3}}\rangle |S_{b}S_{b_{3}}\rangle  |L\ L_{3}\rangle$ for fixed values of $ S_{\mathcal{D}} $, $ S_{b} $ and $ L $. Following that, Eq. [\ref{B1}] is rewritten as 
 		\begin{equation}\label{B2}
 			|^{2S+1}L_{J};J_{3}\rangle = \displaystyle\sum_{\scriptscriptstyle {S_{\mathcal{D}_{3}}S_{b_{3}}L_{3}}S_{3}} C_{S_{\mathcal{D}_{3}}S_{b_{3}}S_{3}}^{S_{\mathcal{D}}\ S_{b}\ S}\ C_{S_{3}L_{3}J_{3}}^{S\ L\ J}\ |S_{\mathcal{D}_{3}},S_{b_{3}},L_{3}\rangle.  
 		\end{equation}
 		
 		Based on the relation above, we can list the  $L-S$  coupling basis states for $F$-wave as follows:
 		
 		\begin{equation} \label{B3}
 			|^{4}F_{3/2};3/2\rangle =-\frac{2}{\sqrt{7}}|-1,-\frac{1}{2},3\rangle+\frac{2}{\sqrt{21}}|0,-\frac{1}{2},2\rangle-\frac{2}{\sqrt{105}}|1,-\frac{1}{2},1\rangle+\sqrt{\frac{2}{21}}|-1,\frac{1}{2},2\rangle-2 \sqrt{\frac{2}{105}}|0,\frac{1}{2},1\rangle+\frac{1}{\sqrt{35}}|1,\frac{1}{2},0\rangle ,
 		\end{equation}
 		\begin{equation} \label{B4}
 			|^{2}F_{5/2};5/2\rangle =-\sqrt{\frac{2}{7}}|0,-\frac{1}{2},3\rangle+\sqrt{\frac{2}{21}}|1,-\frac{1}{2},2\rangle+\frac{2}{\sqrt{7}}|-1,\frac{1}{2},3\rangle-\frac{1}{\sqrt{21}}|0,\frac{1}{2},2\rangle ,
 		\end{equation}
 		\begin{equation} \label{B5}
 			|^{4}F_{5/2};5/2\rangle = \sqrt{\frac{5}{14}}|0,-\frac{1}{2},3\rangle-\sqrt{\frac{5}{42}}|1,-\frac{1}{2},2\rangle+\frac{1}{2}\sqrt{\frac{5}{7}}|-1,\frac{1}{2},3\rangle-\sqrt{\frac{5}{21}}|0,\frac{1}{2},2\rangle+\frac{1}{2}\sqrt{\frac{3}{7}}|1,\frac{1}{2},1\rangle,
 		\end{equation}
 		\begin{equation} \label{B6}
 			|^{2}F_{7/2};7/2\rangle =\sqrt{\frac{2}{3}}|1,-\frac{1}{2},3\rangle-\frac{1}{\sqrt{3}}|0,\frac{1}{2},3\rangle ,
 		\end{equation}
 		\begin{equation} \label{B7}
 			|^{4}F_{7/2};7/2\rangle =-\frac{\sqrt{2}}{3}|1,-\frac{1}{2},3\rangle-\frac{2}{3}|0,\frac{1}{2},3\rangle+\frac{1}{\sqrt{3}}|1,\frac{1}{2},2\rangle ,
 		\end{equation}
 		\begin{equation} \label{B8}
 			|^{4}F_{9/2};9/2\rangle =|1,\frac{1}{2},3\rangle ,
 		\end{equation}
 		
 		We can simplify the operators that are involved in spin-dependent interactions as below:
 		\begin{equation}
 			\mathbf{L\cdot S_{i}}=\frac{1}{2}\left[L_{+}S_{i-}+L_{-}S_{i+}\right]+L_{3}S_{i3},
 		\end{equation}
 		where $i = \mathcal{D}$ or $b$, and
 		\\
 		
 		\begin{equation}
 			 \mathbf{\hat{B}} =\frac{-3}{(2L-1)(2L+3)}\left[(\mathbf{L\cdot S_{\mathcal{D}}})(\mathbf{L\cdot S_{b}})+(\mathbf{L\cdot S_{b}})(\mathbf{L\cdot S_{\mathcal{D}}})-\frac{2}{3}L(L+1)(\mathbf{S_{\mathcal{D}}\cdot S_{b}})\right].
 		\end{equation}
 		The expression for the expectation value of $\mathbf{S_\mathcal{D}}\cdot \mathbf{S_b}$ in  $L-S$  coupling basis is 
 		
 		\begin{equation}
 			\langle\mathbf{S_\mathcal{D}}\cdot \mathbf{S_b}\rangle=\frac{1}{2}[S(S+1)-S_\mathcal{D}(S_\mathcal{D}+1)-S_b(S_b+1)].
 		\end{equation}
 		
 		Then, we find the expectation values of spin-dependent operators in $[^{2}F_{J}, ^{4}F_{J}]$ basis for different J values, and the outcomes are given below:

 		For J=3/2,
 		\begin{equation}
 			\text{$\langle\mathbf{L}\cdot \mathbf{S_\mathcal{D}}\rangle$ =}-4	,\ \text{$\langle\mathbf{L}\cdot \mathbf{S_b}\rangle$ =}-2	,\ \text{$\langle \mathbf{\hat{B}} \rangle$=}-\frac{4}{5}		,\ \text{$\langle\mathbf{S_\mathcal{D}}\cdot \mathbf{S_b}\rangle$=}\frac{1}{2}	.
 		\end{equation}
 		
 		For J=5/2,
 		\begin{equation}
 			\text{$\langle\mathbf{L}\cdot \mathbf{S_\mathcal{D}}\rangle$=}\left[
 			\begin{array}{cc}
 				-\frac{8}{3} & -\frac{2 \sqrt{5}}{3} \\[5pt]
 				-\frac{2 \sqrt{5}}{3} & -\frac{7}{3} \\[1pt]
 			\end{array}		
 			\right],\ \   \text{$\langle\mathbf{L}\cdot \mathbf{S_b}\rangle$=}\left[
 			\begin{array}{cc}
 				\frac{2}{3} & \frac{2 \sqrt{5}}{3} \\[5pt]
 				\frac{2 \sqrt{5}}{3} & -\frac{7}{6} \\[1pt]
 			\end{array}
 			\right],\ \   \text{$\langle \mathbf{\hat{B}} \rangle$=}\left[
 			\begin{array}{cc}
 				0 & \frac{1}{\sqrt{5}} \\[5pt]
 				\frac{1}{\sqrt{5}} & \frac{1}{5} \\[1pt]
 			\end{array}
 			\right],\ \ 
 			\text{$\langle\mathbf{S_\mathcal{D}}\cdot \mathbf{S_b}\rangle$=}\left[
 			\begin{array}{cc}
 				-1 & 0 \\[5pt]
 				0 & \frac{1}{2} \\[1pt]
 			\end{array}
 			\right].
 		\end{equation}
 		
 		For J=7/2,
 		\begin{equation}
 			\text{$\langle\mathbf{L}\cdot \mathbf{S_\mathcal{D}}\rangle$=}\left[
 			\begin{array}{cc}
 				2 & -\sqrt{3} \\[5pt]
 				-\sqrt{3} & 0 \\[1pt]
 			\end{array}
 			\right],\ \   \text{$\langle\mathbf{L}\cdot \mathbf{S_b}\rangle$=}\left[
 			\begin{array}{cc}
 				-\frac{1}{2} & \sqrt{3} \\[5pt]
 				\sqrt{3} & 0 \\[1pt]
 			\end{array}
 			\right],\ \   \text{$\langle \mathbf{\hat{B}} \rangle$=}\left[
 			\begin{array}{cc}
 				0 & -\frac{1}{2 \sqrt{3}} \\[5pt]
 				-\frac{1}{2 \sqrt{3}} & \frac{2}{3} \\[1pt]
 			\end{array}
 			\right],\ \ 
 			\text{$\langle\mathbf{S_\mathcal{D}}\cdot \mathbf{S_b}\rangle$=}\left[
 			\begin{array}{cc}
 				-1 & 0 \\[5pt]
 				0 & \frac{1}{2} \\[1pt]
 			\end{array}
 			\right].
 		\end{equation}

 		For J=9/2,
 		\begin{equation}
 			\text{$\langle\mathbf{L}\cdot \mathbf{S_\mathcal{D}}\rangle$ =}3	,\ \text{$\langle\mathbf{L}\cdot \mathbf{S_b}\rangle$ =}\frac{3}{2}	,\ \text{$\langle \mathbf{\hat{B}} \rangle$=}	-\frac{1}{3}	,\ \text{$\langle\mathbf{S_\mathcal{D}}\cdot \mathbf{S_b}\rangle$=}\frac{1}{2}	.
 		\end{equation}
 		
 		Here, we observe that the dominant interaction term in spin-dependent interaction, i.e., $\langle\mathbf{L}\cdot \mathbf{S_\mathcal{D}}\rangle$, is not diagonal for $J=5/2$ and $J=7/2$, in   $[^{2}F_{J}, ^{4}F_{J}]$ basis of the  $L-S$  coupling scheme. 
 		
 		But in the  $j-j$  coupling scheme, they are diagonal.  To find expectation values of these operators in the  $j-j$  coupling scheme, we start with finding eigen functions corresponding to each eigen value $k$ of $\langle\mathbf{L}\cdot \mathbf{S_\mathcal{D}}\rangle$ which forms the basis in  $j-j$  coupling scheme, as shown below:
 		\begin{equation}
 			|J= \frac{3}{2}	, j=2 \rangle =|^{4}F_{3/2}\rangle
 		\end{equation}
 		\begin{equation}
 			 k =-4	:|J=\frac{5}{2} 	, j=2	 \rangle =\frac{\sqrt{5}}{3}|^{2}F_{5/2}\rangle+\frac{2}{3}|^{4}F_{5/2}\rangle,
 		\end{equation}
 		\begin{equation}
 			k =-1	:|J=\frac{5}{2}	, j=3	 \rangle =-\frac{2}{3}|^{2}F_{5/2}\rangle+\frac{\sqrt{5}}{3}|^{4}F_{5/2}\rangle,
 		\end{equation}
 		\begin{equation}
 			k =3	:|J= \frac{7}{2} 	, j=3	 \rangle =-\frac{\sqrt{3}}{2} 		 |^{2}F_{	7/2}\rangle+\frac{1}{2}		|^{4}F_{	7/2}\rangle,
 		\end{equation}
 		\begin{equation}
 			k =-1	:|J= \frac{7}{2} 	, j=4	 \rangle =\frac{1}{2} 		 |^{2}F_{	7/2}\rangle+\frac{\sqrt{3}}{2}		|^{4}F_{	7/2}\rangle,
 		\end{equation}
 		\begin{equation}
 			|J=\frac{9}{2} , j=4 \rangle =|^{4}F_{9/2}\rangle
 		\end{equation}
 		Following that, we compute the expectation value of spin-dependent operators in a $|J, j \rangle $ basis and display the results in a Table  \ref{tab:table1}.
 		
 		\item {\textit{The G-wave}}: We start with forming the  $L-S$  coupling states as a linear combination of uncoupled states $|S_{\mathcal{D}_{3}},S_{b_{3}}, L_{3}\rangle$, using Eq.(\ref{B2}), as follows: 
 		\begin{equation} \label{}
 			|^{4}G_{5/2};5/2\rangle =-\sqrt{\frac{2}{3}}|-1,-\frac{1}{2},4\rangle+\frac{1}{\sqrt{6}}|0,-\frac{1}{2},3\rangle-\frac{1}{\sqrt{42}}|1,-\frac{1}{2},2\rangle+\frac{1}{2 \sqrt{3}}|-1,\frac{1}{2},3\rangle-\frac{1}{\sqrt{21}}|0,\frac{1}{2},2\rangle+\frac{1}{2 \sqrt{21}}|1,\frac{1}{2},1\rangle ,
 		\end{equation}
 		\begin{equation} \label{}
 			|^{2}G_{7/2};7/2\rangle =-\frac{2}{3}\sqrt{\frac{2}{3}}|0,-\frac{1}{2},4\rangle+\frac{1}{3}\sqrt{\frac{2}{3}}|1,-\frac{1}{2},3\rangle+\frac{4}{3 \sqrt{3}}|-1,\frac{1}{2},4\rangle-\frac{1}{3 \sqrt{3}}|0,\frac{1}{2},3\rangle ,
 		\end{equation}
 		\begin{equation} \label{}
 			|^{4}G_{7/2};7/2\rangle =\frac{2}{3}\sqrt{\frac{14}{15}}|0,-\frac{1}{2},4\rangle-\frac{1}{3}\sqrt{\frac{14}{15}}|1,-\frac{1}{2},3\rangle+\frac{2}{3}\sqrt{\frac{7}{15}}|-1,\frac{1}{2},4\rangle-\frac{2}{3}\sqrt{\frac{7}{15}}|0,\frac{1}{2},3\rangle+\frac{1}{\sqrt{15}}|1,\frac{1}{2},2\rangle ,
 		\end{equation}
 		\begin{equation} \label{}
 			|^{2}G_{9/2};9/2\rangle =\sqrt{\frac{2}{3}}|1,-\frac{1}{2},4\rangle-\frac{1}{\sqrt{3}}|0,\frac{1}{2},4\rangle ,
 		\end{equation}
 		\begin{equation} \label{}
 			|^{4}G_{9/2};9/2\rangle = -2 \sqrt{\frac{2}{33}}|1,-\frac{1}{2},4\rangle-\frac{4}{\sqrt{33}}|0,\frac{1}{2},4\rangle+\sqrt{\frac{3}{11}}|1,\frac{1}{2},3\rangle,
 		\end{equation}
 		\begin{equation} \label{}
 			|^{4}G_{11/2};11/2\rangle =|1,\frac{1}{2},4\rangle ,
 		\end{equation}
 		Following that, the expectation values of spin-dependent operators in $[^{2}G_{J}, ^{4}G_{J}]$ basis are calculated for different values of $J$  and the results are listed below:
 		For J=5/2,
 		\begin{equation}
 			\text{$\langle\mathbf{L}\cdot \mathbf{S_\mathcal{D}}\rangle$ =}-5	,\ \text{$\langle\mathbf{L}\cdot \mathbf{S_b}\rangle$ =}-\frac{5}{2}	,\ \text{$\langle \mathbf{\hat{B}} \rangle$=}-\frac{5}{7}		,\ \text{$\langle\mathbf{S_\mathcal{D}}\cdot \mathbf{S_b}\rangle$=}\frac{1}{2}	.
 		\end{equation}
 		
 		For J=7/2,
 		\begin{equation}
 			\text{$\langle\mathbf{L}\cdot \mathbf{S_\mathcal{D}}\rangle$=}\left[
 			\begin{array}{cc}
 				-\frac{10}{3} & -\frac{\sqrt{35}}{3} \\[5pt]
 				-\frac{\sqrt{35}}{3} & -\frac{8}{3} \\[1pt]
 			\end{array}
 			\right],\ \   \text{$\langle\mathbf{L}\cdot \mathbf{S_b}\rangle$=}\left[
 			\begin{array}{cc}
 				\frac{5}{6} & \frac{\sqrt{35}}{3} \\[5pt]
 				\frac{\sqrt{35}}{3} & -\frac{4}{3} \\[1pt]
 			\end{array}
 			\right],\ \   \text{$\langle \mathbf{\hat{B}} \rangle$=}\left[
 			\begin{array}{cc}
 				0 & \frac{1}{2}\sqrt{\frac{5}{7}} \\[5pt]
 				\frac{1}{2}\sqrt{\frac{5}{7}}  & \frac{2}{7} \\[1pt]
 			\end{array}
 			\right],\ \ 
 			\text{$\langle\mathbf{S_\mathcal{D}}\cdot \mathbf{S_b}\rangle$=}\left[
 			\begin{array}{cc}
 				-1 & 0 \\[5pt]
 				0 & \frac{1}{2} \\[1pt]
 			\end{array}
 			\right].
 		\end{equation}

 		For J=9/2,
 		\begin{equation}
 			\text{$\langle\mathbf{L}\cdot \mathbf{S_\mathcal{D}}\rangle$=}\left[
 			\begin{array}{cc}
 				\frac{8}{3} & -\frac{2 \sqrt{11}}{3} \\[5pt]
 				-\frac{2 \sqrt{11}}{3} & \frac{1}{3} \\[1pt]
 			\end{array}
 			\right],\ \   \text{$\langle\mathbf{L}\cdot \mathbf{S_b}\rangle$=}\left[
 			\begin{array}{cc}
 				-\frac{2}{3} & \frac{2 \sqrt{11}}{3} \\[5pt]
 				\frac{2 \sqrt{11}}{3} & \frac{1}{6} \\[1pt]
 			\end{array}
 			\right],\ \   \text{$\langle \mathbf{\hat{B}} \rangle$=}\left[
 			\begin{array}{cc}
 				0 & -\frac{1}{\sqrt{11}} \\[5pt]
 				-\frac{1}{\sqrt{11}} & \frac{7}{11} \\[1pt]
 			\end{array}
 			\right],\ \ 
 			\text{$\langle\mathbf{S_\mathcal{D}}\cdot \mathbf{S_b}\rangle$=}\left[
 			\begin{array}{cc}
 				-1 & 0 \\[5pt]
 				0 & \frac{1}{2} \\[1pt]
 			\end{array}
 			\right].
 		\end{equation}
 		
 		For J=11/2,
 		\begin{equation}
 			\text{$\langle\mathbf{L}\cdot \mathbf{S_\mathcal{D}}\rangle$ =}4	,\ \text{$\langle\mathbf{L}\cdot \mathbf{S_b}\rangle$ =}2	,\ \text{$\langle \mathbf{\hat{B}} \rangle$=}	-\frac{4}{11}	,\ \text{$\langle\mathbf{S_\mathcal{D}}\cdot \mathbf{S_b}\rangle$=}\frac{1}{2}	.
 		\end{equation}
 		
 		The basis states for the  $j-j$  coupling scheme are formed by the eigen functions corresponding to each eigen value ($k$) of  $\langle\mathbf{L}.\mathbf{S_\mathcal{D}}\rangle$, which are shown below :
 		
 		\begin{equation}
 			|J=\frac{5}{2},j=3\rangle =|^{4}G_{5/2}\rangle
 		\end{equation}
 		\begin{equation}
 			k =-5:|J=\frac{7}{2},j=3\rangle =\frac{1}{2}\sqrt{\frac{7}{3}}|^{2}G_{7/2}\rangle+\frac{1}{2}\sqrt{\frac{5}{3}}|^{4}G_{7/2}\rangle,
 		\end{equation}
 		\begin{equation}
 			k =-1:|J=\frac{7}{2},j=4\rangle =-\frac{1}{2}\sqrt{\frac{5}{3}}|^{2}G_{7/2}\rangle+\frac{1}{2}\sqrt{\frac{7}{3}}|^{4}G_{7/2}\rangle,
 		\end{equation}
 		\begin{equation}
 			k =4:|J=\frac{9}{2}, j=4\rangle =-\sqrt{\frac{11}{15}}|^{2}G_{9/2}\rangle+\frac{2}{\sqrt{15}}|^{4}G_{9/2}\rangle,
 		\end{equation}
 		\begin{equation}
 			k =-1:|J=\frac{9}{2}, j=5\rangle =\frac{2}{\sqrt{15}}|^{2}G_{9/2}\rangle+\sqrt{\frac{11}{15}}|^{4}G_{9/2}\rangle,
 		\end{equation}
 		\begin{equation}
 			|J=\frac{11}{2},j=5\rangle=|^{4}G_{11/2}\rangle
 		\end{equation}
 		With these basis states on hand, we extract the expectation values of operators involved in interactions in the  $j-j$  coupling scheme, and the results are summarised in Table \ref{tab:table1}.
 		
 	\end{enumerate}
 \end{widetext} 

\bibliography{apssamp}

\end{document}